\documentclass[12pt,preprint]{aastex}

\usepackage{xcolor}

\shorttitle{The interaction between shear and fingering (thermohaline) convection}
\shortauthors{Garaud et al.}

\begin{document}


\title{The interaction between shear and fingering (thermohaline) convection}



\author{P. Garaud$^1$, A. Kumar$^1$ and J. Sridhar$^1$}
\affil{
$^{1}$Department of Applied Mathematics, Baskin School of Engineering, \\
University of California, Santa Cruz, CA 95064\\
}



\begin{abstract}
 Fingering convection is a turbulent mixing process that can occur in stellar radiative regions whenever the mean molecular weight increases with radius. In some cases, it can have a significant observable impact on stellar structure and evolution. The efficiency of mixing by fingering convection as a standalone process has been studied by Brown et al. (2013), but other processes such as rotation, magnetic fields and shear can affect it. In this paper, we present a first study of the effect of shear on fingering (thermohaline) convection in astrophysics. Using Direct Numerical Simulations we find that a moderate amount of shear (that is not intrinsically shear-unstable) always decreases the mixing efficiency of fingering convection, as a result of the tilt it imparts to the fingering structures. We propose a simple analytical extension of the Brown et al. (2013) model in the presence of shear that satisfactorily explains the numerically-derived turbulent compositional mixing coefficient for moderate shearing rates, and can trivially be implemented in stellar evolution codes. We also measure from the numerical simulations a turbulent viscosity, and find that the latter is strongly tied to the turbulent compositional mixing coefficient. Observational implications and caveats of the model are discussed. 
\end{abstract}



\section{Introduction}
\label{sec:intro}

The double-diffusive fingering instability (also called thermohaline instability) can be a significant source of turbulent mixing in radiative regions of stars
where the mean molecular weight $\mu$ increases with radius \citep{Ulrich1972}. These so-called inverse $\mu$ gradients can arise in different situations \citep[see the review by][]{Garaud18}. 
For instance, they form whenever a star accretes high $\mu$ material on its surface, either from planetary infall \citep{vauclair2004mfa,Garaud2011,TheadoVauclair2012}, planetary debris accretion and especially on White Dwarfs \citep{Deal2013,Wachlinal2017,BauerBildsten2018}, or mass transfer in binaries \citep{MarksSarna1998,ChenHan2004,Stancliffe2007}. They can also form as a result of the radiative levitation of high-opacity atomic species such as iron and nickel \citep{Theadoal2009,Zemskova2014,Dealal2016}, or from nuclear shell-burning in reactions that decrease the mean molecular weight \citep{Ulrich1972,CharbonnelZahn2007,DenissenkovMerryfield2011}, such as $^3$He + $^3$He $\rightarrow$ p + p + $^4$He.

Double-diffusive instabilities in general were first discussed in the oceanographic context by \citet{Stommel1956} and \citet{stern1960sfa}. They occur in fluids whose overall density profile is stable to standard convection (i.e. Ledoux-stable in the stellar context), but where the density itself depends on multiple components that diffuse at different rates. When the rapidly-diffusive component (here, temperature) is stably stratified while the slowly diffusive component (here, the chemical composition) is unstably stratified, as it would be in stellar radiative zones with an inverse $\mu$ gradient, the fingering instability can take place. Fluid parcels displaced from high $\mu$ regions downward rapidly equilibrate thermally with their surroundings (due to the fast thermal diffusion) but retain most of their excess $\mu$ (due to the slow compositional diffusion). As a result, they become denser than the environment and sink further. The nonlinear development of the instability in stellar interiors takes the form of roughly spherical parcels of higher entropy, higher $\mu$ fluid moving down, and lower entropy, lower $\mu$ fluid moving up \citep{Traxler2011b}.  

\citet{Ulrich1972} and \citet{kippenhahn80} were the first to propose a parametrization for mixing by the fingering instability in the context of stellar astrophysics. While their approaches differ substantially from one another, the mixing coefficient they arrive at is the same aside from a constant multiplicative factor $\xi$: 
\begin{equation}
\kappa_{\rm turb} = \xi \frac{ \kappa_T}{R_0-1}  \quad ,
\label{eq:UKRT}
\end{equation}
where $\kappa_T$ is the thermal diffusivity, $R_0 = \delta (\nabla - \nabla_{\rm ad}) / \phi \nabla_\mu$ is the density ratio, and where $\delta$, $\phi$, $\nabla$, $\nabla_{\rm ad}$ and $\nabla_\mu$ are defined as usual in stellar evolution \citep[e.g.][]{kippenhahnweigert}. \citet{Ulrich1972} argued that the constant $\xi$ would be of the order of 1000, while \citet{kippenhahn80} put forward a much smaller value of the order of 10. Nowadays this simple model is commonly used in stellar evolution computations, with the constant $\xi$ left as a tunable parameter \citep[as in][for instance]{MESA12011}.

Recent developments in computational fluid dynamics have revitalized the field, and provide means of performing direct numerical simulations (DNSs) to study the nonlinear behavior of the fingering instability in detail. Early works along those lines \citep{denissenkov2010,Traxler2011b,DenissenkovMerryfield2011} focussed on testing the validity of the turbulent mixing prescription (\ref{eq:UKRT}), and unanimously ruled in favor of the smaller constant proposed by \citet{kippenhahn80}. The more systematic numerical experiments of \citet{Traxler2011b} and \citet{Brownal2013} however revealed an additional dependence of the mixing coefficient on the Schmidt number (i.e. the ratio of the kinematic viscosity $\nu$ to the compositional diffusivity $\kappa_C$), that had so far been ignored. To explain their results, \citet{Brownal2013} \citep[see also][]{RadkoSmith2012} proposed that the exponential growth of the fingering instability stalls when parasitic shear instabilities between adjacent fingers cause their nonlinear saturation. They assumed that this happens when the growth rate of these parasitic instabilities is of the order of the growth rate of the basic fingering instability (see Section \ref{sec:simplemodel} for more detail). Using this idea, they were able to quantify the nonlinear saturation of the fingering instability and to better account for the variation of the turbulent mixing coefficient with input parameters measured in the DNSs.  

More recently, \citet{Garaudal2015} looked into the geophysically-inspired notion that the basic fingering instability might excite larger-scale dynamics such as internal gravity waves or thermo-compositional staircases in stars, but ultimately showed that these processes do not seem to take place at stellar parameters, at least in the absence of any other physical process at play. This naturally raised the follow-up question of whether (and how) processes such as rotation, horizontal temperature and compositional gradients, magnetic fields or shear would affect the dynamics of fingering convection. While most of the subsequent research remains quite preliminary, interesting discoveries have been made. \citet{Medrano2014} \citep[see also][]{Holyer1983} for instance showed that horizontal gradients greatly widen the region of parameter space that is linearly unstable, to the extent that {\it any} inverse $\mu$ gradient, however weak, could drive so-called lateral intrusions. However, the turbulent mixing coefficient in the regime stable to fingering convection but unstable to lateral intrusions remains weak. \citet{SenguptaGaraud2018} considered fingering convection in a rotating fluid, with the assumption that the rotation axis is aligned with gravity (i.e. in the polar regions of a star). They found that rotation generally reduces the efficiency of turbulent mixing, except in some specific regions of parameter space where large-scale vortices aligned with the rotation axis form. The presence of these vortices appears to greatly enhance vertical mixing by fingering convection, by many orders of magnitude above what one would observe in the non-rotating case. Whether such vortices are actually present in stars, however, is unknown. Finally, \citet{HarringtonGaraud2019} looked at the effect of large-scale magnetic fields. They found that when the field is aligned with gravity (and consequently with the fastest-growing fingering modes), it stabilizes the fingers against their parasitic instabilities and allows them to grow to much larger amplitudes before saturating. This greatly enhances transport by fingering convection compared with the non-magnetic case.

An important process that remains to be investigated is the effect of shear on fingering instabilities. Shear is always present in stars, albeit with a wide range of amplitudes, lengthscales and timescales. Global-scale slowly varying shear flows are observed in stars thanks to helio- and asteroseismology. For instance the solar tachocline is a thin rotational shear layer located at the base of the convection zone, whose radial shear is about 10\% of the mean rotation rate of the Sun  \citep{JCDSchou88,Brownal1989,Charbonneaual1999}. The envelopes of subgiants are observed to rotate significantly slower than their cores, demonstrating the presence of (presumably) large-scale rotational shear as well \citep{Beck2012,Deheuvels2014}. Smaller-scale and more rapidly varying shear flows are also very likely in stars. They could arise for instance from the interaction between gravity waves and rotation (in a manner similar to the quasi-biennal oscillation in the Earth's atmosphere), as discussed by \citet{Kumaral1999} in the solar context. They could also be driven by tides in close binary stars \citep[see the review by][]{Ogilvie2014}, as well as a large number of other possible mechanisms. Since the size of fingering structures is typically of the order of meters to hundreds of meters at stellar parameters, even a small-scale shear flow from the perspective of the entire star would appear to be large-scale from the perspective of the fingers themselves. We can therefore ask the question of how a ``large-scale" shear affects fingering convection in stars, and in particular, its associated turbulent mixing properties. 

The related question of how fingering convection transports momentum and influences shear flows is also of interest, especially in the context of the aforementioned observations of rotational shear in subgiant stars. Indeed, these observations are at odds with predictions from stellar evolution calculations in which angular momentum is conserved by each spherical mass shell as the core contracts and the envelope expands. Such calculations predict a much higher level of differential rotation than what is observed \citep{Marques2013}. To reconcile theory and observations requires efficient angular momentum transport between the core and the envelope as the star enters and evolves along the subgiant branch. Fingering convection has naturally been proposed as a possible mechanism for turbulent momentum transport in these objects, but little is known about its efficiency. The latter is usually assumed to be proportional to the turbulence compositional mixing coefficient, with the proportionality constant left as a tunable parameter again \citep[cf.][]{MESA12011}. Whether this prescription is reasonable or not remains to be determined.
 
In what follows we therefore undertake a preliminary set of numerical experiments to quantify mixing of both chemical species and momentum in sheared fingering convection in stellar interiors. To our knowledge, this has never been studied before, though related work on the effect of shear on geophysical fingering convection exists \citep[][see below for a discussion of their results]{Linden1974,ruddick1985,kunze1990,kunze1994,wells2001,SmythKimura2007,KimuraSmyth2007,Radkoal2015}. Section \ref{sec:model} presents our mathematical model, and Section \ref{sec:linal} analyses its linear stability properties. In Section \ref{sec:num} we then describe and analyse a series of DNSs spanning a wide region of parameter space and discuss the emergence of two distinct regimes: a fingering-dominated regime, and a shear-dominated regime. Section \ref{sec:simplemodel} presents a very simple geometric model to explain the various quantitative trends observed in our simulations, and Section \ref{sec:ccl} concludes by discussing astrophysical implications of our results, and future work.

\section{Model setup}
\label{sec:model}

Given that our goal is to study the effect of shear on fingering instabilities in the context of stellar interiors where there are no solid boundaries, we must adopt a model setup that is also free of solid boundaries. To do so, it is common to use a computational domain that is periodic in all directions. In order to add shear to this triply-periodic system, there are two possibilities: the first is to use a shearing box setup, which assumes the existence of a fixed uniform shear throughout the domain, and the second is a body-forced setup, in which the shear flow is driven by a steady spatially-periodic force. We adopt the body-forced setup primarily because it enables us to study the two-way interaction between fingering and shear, and in particular, how the turbulent flux of momentum induced by fingering convection controls the evolution of the large-scale shear flow (which cannot be done in the shearing box model). As we demonstrate in Section \ref{sec:simplemodel}, however, there are also drawbacks to this setup, so further work using a shearing box will be presented in a follow-up publication.

In all that follows, we model a small region of a star using a local Cartesian domain $(x,y,z)$ with gravity aligned with the vertical axis: ${\bf g} = -g{\bf e}_z$. We ignore the effects of rotation and magnetic fields for now, and refer the reader to the works of \citet{SenguptaGaraud2018} and \citet{HarringtonGaraud2019} for preliminary results on the topic. {\bf We use the Boussinesq approximation for weakly compressible gases \citep{SpiegelVeronis1960}, which is valid as long as 
the height of the computational domain $L_z$ is much smaller than any density or temperature scaleheight, and the typical turbulent velocities are much smaller than the local sound speed. Both conditions are satisfied deep in stellar interiors, primarily because double-diffusive turbulence at stellar parameters is both weak and small scale \citep{Garaud18}. However, they may fail very close to the surface where both the pressure scaleheight and the sound speed drop significantly, and the double-diffusive lengthscale increases. As such, we restrict the scope of this study to double-diffusive regions that are well below the surface. Consistent with the Boussinesq approximation,} we also take $g$ and all the diffusivities ($\nu$, $\kappa_T$ and $\kappa_C$) to be constant within the domain.

We then write the temperature and composition as the sum of a linear background stratification plus perturbations that are triply periodic: 
\begin{eqnarray}
T(x,y,z,t) & = & T_0(z) + \tilde{T}(x,y,z,t), \\
C(x,y,z,t) & = & C_0(z) + \tilde{C}(x,y,z,t),
\end{eqnarray}
where $T_0(z)=T_m+T_{0z}z$ and $C_0(z)=C_m + C_{0z}z$, where $T_m$, $C_m$, $T_{0z}$ and $C_{0z}$ are constant. The compositional field $C$ can be viewed as the concentration of a particular species per unit mass, or as the mean molecular weight $\mu$. As required by the Boussinesq approximation, we linearize the equation of state around the mean state denoted by the subscript $m$ so the mean density profile $\rho_0(z)$ and perturbations $\tilde{\rho}$ are given by 
\begin{eqnarray}
\frac{\rho_0(z)}{\rho_m} = 1 - \alpha T_{0z} z + \beta C_{0z} z \\ 
\frac{\tilde{\rho}}{\rho_m} = -\alpha \tilde{T} + \beta \tilde{C}, \label{eq:eos}
\end{eqnarray}
where $\rho_m = \rho(p_m,T_m,C_m)$ is the mean density of the region, $p_m$ is the mean pressure, and where $\alpha= -\rho_m^{-1} \frac{\partial \rho}{\partial T}$, and $\beta= \rho_m^{-1} \frac{\partial \rho}{\partial C}$ are the corresponding partial derivatives of the equation of state at $T_m$, $C_m$, and $p_m$. 

The Boussinesq equations governing the fluid evolution under the effect of a body-force ${\bf F}$ are:
\begin{eqnarray}
\frac{\partial{\bf u}}{\partial t} + {\bf u}\cdot\nabla {\bf u} & = & - \frac{1}{\rho_m} \nabla \tilde{p} + (\alpha \tilde{T} -\beta  \tilde{C}) g {\bf e}_z + \nu \nabla^2 {\bf u} +  \frac{1}{\rho_m} {\bf F} \label{eq:momentum}, \\
\nabla \cdot {\bf {u}} &=& 0 \label{eq:continuity},  \\
\frac{\partial \tilde{T}}{\partial t} + {\bf u}\cdot\nabla \tilde{T} + w  (T_{0z} - T_{{\rm ad},z})  & = & \kappa_T \nabla^2 \tilde{T} \label{eq:heat}, \\
\frac{\partial \tilde{C}}{\partial t} + {\bf u}\cdot\nabla \tilde{C} + w C_{0z} & = & \kappa_C \nabla^2 \tilde{C} \label{eq:composition},
\end{eqnarray}
where ${\bf u} = (u,v,w)$, $T_{{\rm ad},z} = - g/c_p$ is the adiabatic temperature gradient, and $c_p$ is the specific heat at constant pressure. 

In all that follows, the shear will be driven by a sinusoidal body force ${\bf F} = F_0 \sin(k_s z) {\bf e}_x$ where $k_s  = 2\pi/L_z$ is the wavenumber associated with the domain height. With this assumption, the laminar steady state solution of the governing equations is 
\begin{equation}
{\bf u}_L(x,y,z) = \frac{F_0}{\rho_m \nu k_s^2} \sin(k_s z) {\bf e}_x .
\end{equation}
For this laminar flow, the maximum value of the shear is $|S_L | = \frac{F_0}{\rho_m \nu k_s}$, and is achieved at $z = 0$, $z = L_z / 2$, and $z = L_z$. 
The constant buoyancy frequency in this system is 
\begin{equation}
N^2  \equiv N_T^2 + N_C^2 = \alpha g (T_{0z} - T_{{\rm ad},z})  - \beta g C_{0z}, 
\end{equation}
where $N_T^2 = \alpha g (T_{0z} - T_{{\rm ad},z})$ and $N_C^2 = - \beta g C_{0z}$ are the contributions from the temperature and compositional stratifications, respectively.
With this information we can compute the Richardson number of the laminar flow as
\begin{equation}
{\rm Ri} = \frac{N^2}{S_L^2} = ( \alpha g (T_{0z} -  T_{{\rm ad},z})  - \beta g C_{0z}) \left(\frac{\rho_m \nu k_s}{F_0} \right)^2 .
\label{eq:Ri1}
\end{equation}

It is customary in studies of double-diffusive convection to non-dimensionalize the variables and equations using the anticipated finger scale,  
\begin{equation}
d = \left( \frac{ \kappa_T \nu}{ N_T^2 } \right)^{1/4},
\end{equation}
as the unit length, the thermal diffusion timescale across $d$ (namely $d^2/\kappa_T$) as the unit time, $d (T_{0z} -  T_{{\rm ad},z}) $ as the unit temperature and $d \alpha (T_{0z} -  T_{{\rm ad},z})/\beta$ as the unit composition. The non-dimensional equations are then
\begin{eqnarray}
\frac{\partial \hat{\bf u}}{\partial t} + \hat{\bf u}\cdot\nabla \hat{\bf u} & = & - \nabla \hat{p} + {\rm Pr}(\hat{T} -\hat{C}) {\bf e}_z +{\rm Pr} \nabla^2 \hat {\bf u} +  \hat {F_0} \sin(\hat k_s z) {\bf e}_x \label{eq:nondimmomentum}, \\
\nabla \cdot \hat{\bf u} &=& 0 \label{eq:nondimcontinuity},  \\
\frac{\partial \hat{T}}{\partial t} + \hat {\bf u}\cdot\nabla \hat{T} + \hat{w}  & = & \nabla^2 \hat{T} \label{eq:nondimheat}, \\
\frac{\partial \hat{C}}{\partial t} + \hat {\bf u}\cdot\nabla \hat{C} + R_0^{-1} \hat w  & = & \tau \nabla^2 \hat{C} \label{eq:nondimcomposition},
\end{eqnarray}
where all the hatted quantities are from here on non-dimensional\footnote{To simplify the notation, we have not added hats on the independent variables $x,y,z$ and $t$, or on the differential operators; their non-dimensionalization is implicit.} and 
\begin{eqnarray}
{\rm Pr} = \frac{\nu}{\kappa_T}, \quad \tau = \frac{\kappa_C}{\kappa_T}, \\
R_0 = \frac{\alpha (T_{0z} -  T_{{\rm ad},z})}{\beta C_{0z}}, \quad \hat k_s = d k_s  = \frac{2\pi d}{L_z} , \quad \hat F_0 = \frac{F_0}{\rho_m} \frac{d^3}{\kappa_T^2}  .
\end{eqnarray}
The Prandtl number ${\rm Pr}$ is typically very small in stellar interiors, ranging from values of the order of $10^{-2}$ in degenerate regions of White Dwarfs, to values of the order of $10^{-6}$ or even smaller in non-degenerate regions of Red Giants and Main Sequence stars. The diffusivity ratio $\tau$ is also very small in stars, and is typically a few times to an order of magnitude smaller than the Prandtl number. The density ratio $R_0$ is the ratio of the background density gradient due to the entropy (or equivalently potential temperature) stratification, to the background density gradient due to the compositional stratification. When no shear is present, a fingering instability developing on linear background profiles of $T$ and $C$ is known to take place for values of $R_0$ in the interval $[1, \tau^{-1}]$ \citep{baines1969} with $R_0 = 1$ corresponding to the Ledoux criterion (so $R_0<1$ regions are convectively unstable), and $R_0 = \tau^{-1}$ corresponding to marginal stability to fingering convection, see \citet{Garaud18} for a review.

The non-dimensional laminar flow and shear amplitudes are, respectively, 
\begin{equation}
\hat {\bf u}_L(x,y,z) =  \hat u_L   \sin(\hat k_s z) {\bf e}_x \mbox{ where } \hat u_L =  \frac{\hat F_0}{{\rm Pr} \hat k_s^2},  \mbox{    and    }    \hat S_L =  \frac{\hat F_0}{{\rm Pr} \hat k_s},
\label{eq:flowdef}
\end{equation} 
so the Richardson number defined in equation (\ref{eq:Ri1}) can also be expressed as 
\begin{equation}
{\rm Ri} = \frac{ {\rm Pr}^3  \hat k_s^2}{\hat F_0^2}    ( 1 - R_0^{-1} )  = \frac{{\rm Pr} ( 1 - R_0^{-1} )  }{\hat S^2_L}  = \frac{\hat N^2}{ \hat S^2_L}.
\label{eq:Ri}
\end{equation}
which defines $\hat N^2$. This expression is used to select the forcing amplitude $\hat F_0$ for each simulation, so that we can achieve a desired laminar Richardson number ${\rm Ri}$ for any given Prandtl number, density ratio and domain size.

 The stability properties of sinusoidal shear flows in the limit of infinite density ratio (i.e. in the absence of compositional stratification) have been studied extensively. When both viscosity and thermal diffusion are neglected (i.e. by artificially removing the Laplacian in the momentum and temperature equations), the shear is linearly unstable provided ${\rm Ri} < 1/4$, and linearly stable if ${\rm Ri} > 1/4$. This is in accordance with the Howard-Miles theorem for non-diffusive shear flows \citep{miles1961,howard1961}. When viscosity is included, the critical Richardson number can increase to a value of 1, thanks to the emergence of a long-wavelength oscillatory viscous instability \citep{balmforthyoung2002,balmforthyoung2005,Garaudal15a}. When thermal diffusion is also included then the stabilizing effects of the temperature stratification are reduced. In that case the critical Richardson number below which instability exists can increase dramatically \citep[see the works of][]{Zahn1974,Zahn92,PratLignieres13,PratLignieres14,Garaudal15a,GaraudKulen16,Pratal2016,GaraudGagnier2017}, and the system further becomes nonlinearly unstable to small-scale perturbations. An empirical criterion for nonlinear instability found independently by \citet{PratLignieres14} and \citet{GaraudGagnier2017} suggests that turbulence can be sustained in a region with local shear $S$ and local buoyancy frequency $N_T$ provided $ J_T {\rm Pr}  < (J {\rm Pr})_c  \simeq 0.007$, where $J_T = N_T^2/S^2$  is the gradient Richardson number associated with the temperature stratification. 
 
 By contrast, the effect of shear on a double-diffusively stratified system has been given less attention in general, and so far none in the astrophysical context. Theoretical and experimental studies in the geophysical context, i.e. with Prandtl number of order unity, have shown that shear can stabilize fluid motions that vary along the direction of the mean flow (hereafter called the streamwise direction),  but does not affect fluid motions that are streamwise invariant. As a result, so-called salt {\it sheets} are expected instead of salt fingers, and have indeed been observed experimentally \citep{Linden1974,kunze1990,kunze1994,wells2001,SmythKimura2007,KimuraSmyth2007,SmythKimura2010,Radkoal2015}. As summarized by \citet{SmythKimura2010}, these salt-sheets are quite resilient to the presence of a moderate steady shear. The main effect of the shear is to decrease the vertical compositional transport despite the fact that kinetic energy is added into the system. Whether this result continues to hold at low Prandtl numbers appropriate for astrophysical environments is one of the questions we aim to address in this paper. 

In the following section, we now evaluate the stability of the doubly-stratified sinusoidal shear flow $\hat {\bf u}_L$ described above (see equation \ref{eq:flowdef}) to infinitesimal perturbations, and then study its nonlinear evolution in Section \ref{sec:num}.

\section{Linear stability analysis} 
\label{sec:linal}

\subsection{Method} 

Assuming that $\hat {\bf u} = \hat {\bf u}_L + \tilde{\bf u}$, where $|  \tilde{\bf u} | \ll | \hat {\bf u}_L|$, and assuming that $\hat T \equiv \tilde T \ll 1$ (and similarly for the compositional perturbations), we can linearize the governing equations (\ref{eq:nondimmomentum})-(\ref{eq:nondimcomposition}) as 
\begin{eqnarray}
\frac{\partial \tilde {\bf u}}{\partial t} + \hat{\bf u}_L \cdot\nabla \tilde{\bf u} + \tilde{\bf u} \cdot\nabla \hat{\bf u}_L & = & - \nabla \tilde{p} + {\rm Pr}(\tilde{T} -\tilde{C}) {\bf e}_z +{\rm Pr} \nabla^2 \tilde {\bf u}  \label{eq:linearmomentum}, \\
\nabla \cdot \tilde {\bf u} &=& 0 \label{eq:linearcontinuity},  \\
\frac{\partial \tilde{T}}{\partial t} + \hat {\bf u}_L \cdot\nabla \tilde{T} + \tilde {w}  & = & \nabla^2 \tilde{T} \label{eq:linearheat}, \\
\frac{\partial \tilde{C}}{\partial t} + \hat {\bf u}_L \cdot\nabla \tilde{C} + R_0^{-1} \tilde w  & = & \tau \nabla^2 \tilde{C} \label{eq:linearcomposition}.
\end{eqnarray}
This is a set of linear partial differential equations with coefficients that are constant in time and in the horizontal coordinates, but that vary sinusoidally with height $z$ (via $\hat {\bf u}_L $). 

If we consider perturbations that are invariant both in the streamwise and vertical directions (so $\partial / \partial x$ and $\partial / \partial z$ are null), then we must also have $\tilde v = 0$ by continuity. The linear stability of these perturbations is identical to that of the corresponding unsheared ones since they satisfy
\begin{eqnarray}
\frac{\partial \tilde w}{\partial t}  & = &  {\rm Pr}(\tilde{T} -\tilde{C}) +{\rm Pr} \nabla^2 \tilde {w}, \label{eq:saltsheetw} \\
\frac{\partial \tilde{T}}{\partial t} + \tilde {w}  & = & \nabla^2 \tilde{T}, \\
\frac{\partial \tilde{C}}{\partial t} + R_0^{-1} \tilde w  & = & \tau \nabla^2 \tilde{C} ,
\end{eqnarray}
The only difference is in the evolution of $\tilde u$, which would merely decay viscously in the unsheared case, but satisfies
\begin{equation}
\frac{\partial \tilde u}{\partial t} +  \tilde{w} \hat u_L \hat k_s \cos(\hat k_s z) =  {\rm Pr} \nabla^2 \tilde {u}  \label{eq:saltsheetu}, \\
\end{equation}
when the background shear is present. Since $\tilde u$ does not enter in the (linear) evolution of $\tilde w$, $\tilde T$ and $\tilde C$, it does not affect the linear stability of the problem. 
The solutions to the system of equations (\ref{eq:saltsheetw})-(\ref{eq:saltsheetu}) are well known and describe the dynamics of the aforementioned salt sheets (see Section \ref{sec:model}, and e.g. \citet{Linden1974}). For Richardson numbers above the threshold for shear instabilities, they are the fastest growing unstable modes.

On the other hand, in the more general situation where $\partial / \partial x$ is non-zero, then the effect of the shear cannot be ignored. To compute the growth rate of such modes, we assume an ansatz of the form
\begin{equation}
\tilde q(x,y,z,t) = \sum_n q_n \exp( i \hat k_x x + i n \hat k_s z + \hat \lambda t),
\end{equation}
for $\tilde q \in \left\{\tilde{\bf u},\tilde T, \tilde C, \tilde p \right\}$. This now implicitly assumes that the perturbations are two-dimensional by being invariant in the spanwise ($y$) direction, and that they are periodic with the same period in the vertical direction as the background flow. These assumptions are made for simplicity. Substituting this ansatz into the system of equations (\ref{eq:linearmomentum})-(\ref{eq:linearcomposition}), and then projecting the result on individual Fourier modes, we obtain a linear system of coupled eigenvalue equations for the coefficients $q_n$, namely
\begin{eqnarray}
\hat \lambda u_n + \frac{\hat u_L \hat k_x}{2} (u_{n-1} - u_{n+1}) + \frac{\hat k_s \hat u_L}{2} (w_{n-1} + w_{n+1})  = - i \hat k_x p_n - {\rm Pr} ( n^2 \hat k_s^2 + \hat k_x^2) u_n , \label{eq:eigenu} \\
\hat \lambda w_n + \frac{\hat u_L \hat k_x}{2} (w_{n-1} - w_{n+1}) = - i n \hat k_s p_n + {\rm Pr} (T_n - C_n) - {\rm Pr} ( n^2 \hat k_s^2 + \hat k_x^2) w_n ,  \\ 
\hat \lambda T_n + \frac{\hat u_L \hat k_x}{2} (T_{n-1} - T_{n+1})  + w_n = - ( n^2 \hat k_s^2 + \hat k_x^2) T_n,  \label{eq:eigenT} \\
\hat \lambda C_n + \frac{\hat u_L \hat k_x}{2} (C_{n-1} - C_{n+1})  + R_0^{-1} w_n = - \tau ( n^2 \hat k_s^2 + \hat k_x^2) C_n , \label{eq:eigenC} \\
\hat k_x u_n + n \hat k_s w_n  = 0 ,
\end{eqnarray}
for any integer index $n$. To solve this numerically, the Fourier expansion needs to be truncated, so $ - N \leq  n \leq N$. The variables $p_n$ and $u_n$ can easily be eliminated, 
 so 
\begin{equation}
\hat \lambda   w_n  + n \hat k_x \hat k^2_s \frac{\hat u_L }{2K_n^2} ((n-2) w_{n-1} - (n+2) w_{n+1}) + \frac{\hat u_L \hat k^3_x}{2K_n^2} (w_{n-1} - w_{n+1}) = \frac{\hat k^2_x }{K_n^2} {\rm Pr} (T_n - C_n)  - {\rm Pr}  K_n^2 w_n \label{eq:eigenw}
\end{equation}
where $K_n^2 = n^2 \hat k_s^2 + \hat k_x^2$. 

With this, the system can be cast in the form of an eigenvalue problem $\bf A \bf v = \hat\lambda \bf v$ where ${\bf v} = (w_{-N},\dots,w_N, T_{-N},\dots, T_N,C_{-N},\dots,C_N)$, and solved numerically using the LAPACK routine DGEEV. For each set of input parameters (${\rm Pr},\tau; {\rm Ri}, R_0; \hat k_s, \hat k_x$), this procedure yields $3 \times (2N+1)$ eigenvalues, from which we select only the one with the largest real part. Finally, we maximize this growth rate over all possible horizontal wavenumbers $\hat k_x$ to find the growth rate of the fastest-growing (or most slowly decaying) mode for a given physical system. Note that it is quite important to verify that the growth rates found using this method are independent of $N$, the truncation order of the system. This has been checked for all the results presented below.

\subsection{Results}
\label{sec:linres}

\begin{figure}
\epsscale{0.6}
\plotone{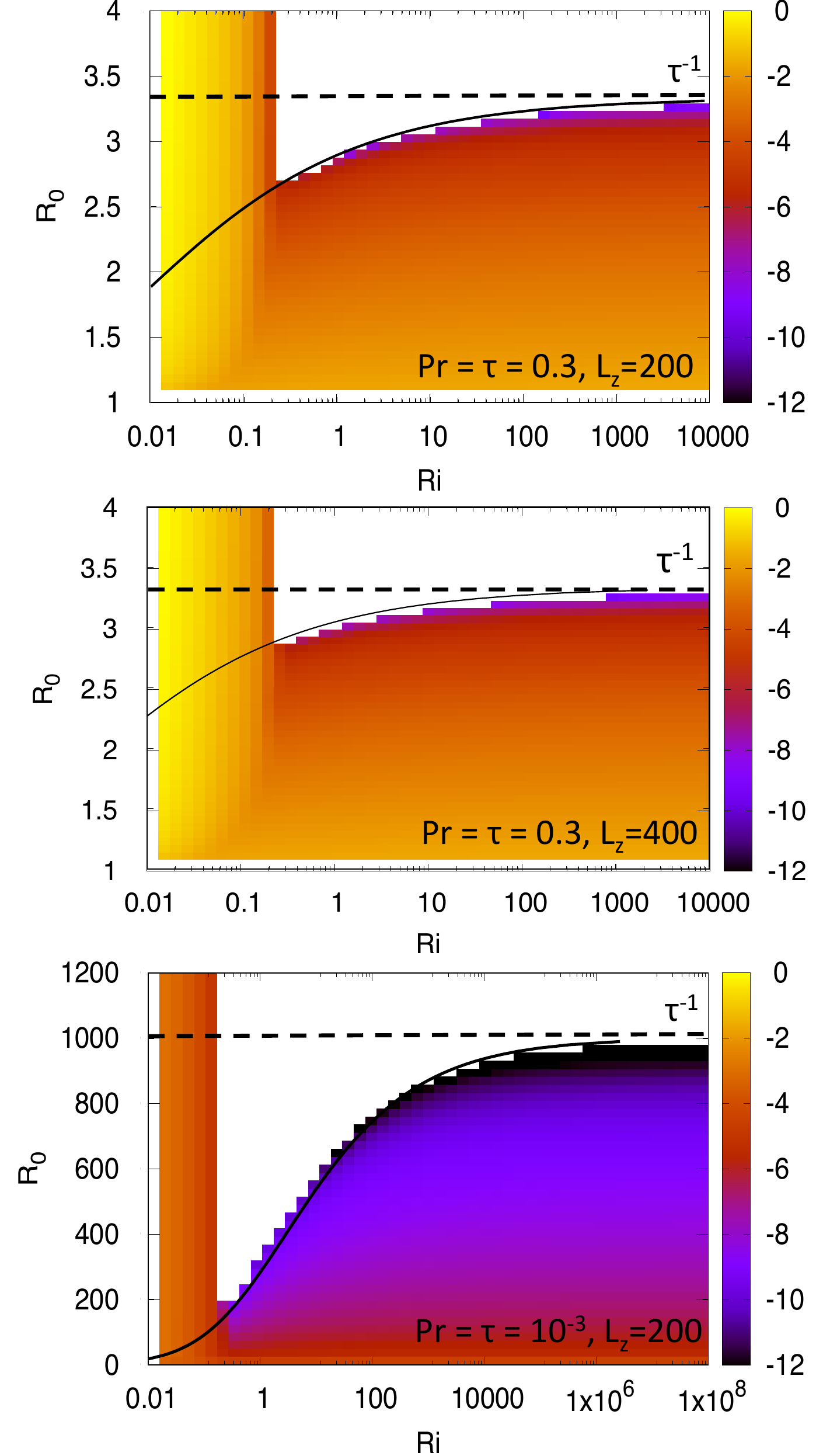}
\caption{Logarithm of the growth rate of the fastest-growing modes of sheared fingering convection, as a function of ${\rm Ri}$ and $R_0$. The values of ${\rm Pr}$, $\tau$ and $\hat k_s$ selected are shown in each panel. Also shown as a vertical dashed line is the stability limit for non-sheared fingering convection ($R_0 = \tau^{-1}$). The curved black line is the level contour $\hat \lambda_f L_z / \hat S_L d = 4$ where $\hat \lambda_f$ is the growth rate of the unsheared fingers. This contour seems to be a good predictor for the edge of the unstable region.}
\label{fig:linstab}
\end{figure}

Figure \ref{fig:linstab} summarizes the linear stability analysis of sheared doubly-stratified systems in the fingering regime to spanwise-invariant periodic perturbations, and presents the results in the form of stability diagrams in the $({\rm Ri},R_0)$ plane for fixed values of ${\rm Pr}$, $\tau$ and $\hat k_s$ (indicated on the bottom right corner). In each panel, the colors represent the logarithm (in base 10) of the growth rate of the fastest-growing mode found using the method described in the previous section, with the eigensystem (\ref{eq:eigenT})-(\ref{eq:eigenC}) and (\ref{eq:eigenw}) truncated at $N = 100$ (we have verified that this sufficiently large for the results to be independent of $N$). A white color is used when all the modes have negative growth rates, implying that the flow is linearly stable. In the top two panels, the Prandtl number Pr and diffusivity ratio $\tau$ are both equal to 0.3. This is  the value used in the majority of the numerical simulations presented in Section \ref{sec:num}. Two values of $\hat k_s$ are presented, for comparison. In the third panel Pr and $\tau$ are both equal to $10^{-3}$, a value that is much lower (though not as low as what one may expect in non-degenerate regions of stellar interiors). These three panels illustrate a number of interesting trends that actually hold for all parameters.

Generally speaking, we find that sheared doubly-stratified systems in the fingering regime can be classified into three categories. For sufficiently small values of the Richardson number ${\rm Ri}$, we find that the system is linearly unstable to shearing modes. These are characterized by a large growth rate commensurate with the flow shearing rate $\hat S_L$, and a small wavenumber $\hat k_x$ of the order of the wavenumber of the imposed shear $\hat k_s$. This mode of instability is relatively independent of the density ratio $R_0$ associated with the thermocompositional stratification. For sufficiently large values of the Richardson number, on the other hand, the effect of the shear on the intrinsic fingering instability is negligible, and the modes of instability are fingering modes, characterized by a horizontal wavenumber of the order of unity (in the non-dimensionalization selected) and a growth rate that depends sensitively on $R_0$, Pr and $\tau$. We recover the well-known result that the system is only unstable in that limit when $R_0 < \tau^{-1}$. 

Finally, for intermediate values of the Richardson number we see that the system is stabilized (to the assumed infinitesimally small $y-$invariant 2D perturbations) for values of the density ratio smaller than $\tau^{-1}$, showing that moderate shear has a stabilizing effect on the system. The shape of the marginal stability curve depends on the values of Pr, $\tau$ and $\hat k_s$ selected. In particular we see that shear has a more strongly stabilizing effect at lower values of Pr and $\tau$ (for fixed $\hat k_s$). To understand qualitatively why this may be the case, recall that the intrinsic growth rate $\hat \lambda_f$  of the pure fingering instability (i.e in the absence of shear) is much smaller at low Prandtl number than at large Prandtl number \citep[e.g.][]{Brownal2013}. As such, the same shearing rate is expected to have a stronger effect on the fingers at low Pr than at high Pr. 

In fact, we found empirically that the ratio $\hat \lambda_f L_z / \hat S_L d $ (where $d= 1$ in the units selected) plays an important role in the linear instability of sheared fingering convection. To see this we have overlaid the level contour $\hat \lambda_f L_z / \hat S_L d = 4$ on the growth rate plots (the constant 4 was fitted to the data). It appears that this contour is a good predictor for the edge of the unstable region (at least approximately) in all cases. To understand the significance of this parameter, note that $\hat \lambda_f L_z / \hat S_L d$ is equal to the ratio of the vertical to horizontal changes in finger velocities (in the linear regime) due to the combined effects of the exponential growth of vertical motions from the fingering instability, and the growth of horizontal motions due to advection by the shear. 
This suggests that shear can stabilize the fingering instability (at least from a linear perspective) as soon as the shearing rate $\hat S_L$ exceeds $(L_z / 4d)  \hat \lambda_f$. 

It is worth remembering, however, that all of these linear stability results were obtained for 2D perturbations that are invariant in $y$, and that streamwise-invariant modes are by contrast unaffected by the shear. In addition, it is also important to note that in stars, ${\rm Ri}$ is unknown since it is the Richardson number associated with a hypothetical laminar shear, rather than the one associated with the actual shear in the system. For both reasons, this linear stability analysis is of limited practical interest, and we now proceed to present the results of nonlinear DNSs of the problem.  

\section{Numerical simulations} 
\label{sec:num}

\subsection{Methodology}

In this section, we use DNSs to investigate the nonlinear evolution of the system until it reaches a statistically stationary state, as well as the properties of that state. We use the pseudo-spectral PADDI code \citep{Traxler2011a,Stellmach2011}, which solves equations (\ref{eq:nondimmomentum})-(\ref{eq:nondimcomposition}) using the Patterson-Orszag algorithm. Linear terms in the equations are advanced implicitly in spectral space, while nonlinear terms are computed first in real space, then transformed back into spectral space before advancing them explicitly using a third order Adams-Bashforth backward differencing algorithm. The code has been extensively used in the astrophysical literature to study both double-diffusive instabilities \citep[e.g.][]{Traxler2011b,rosenblumal2011,Mirouh2012,Brownal2013,Woodal13} and shear instabilities \citep[e.g.][]{Garaudal15a,GaraudKulen16,GagnierGaraud2018}. 

The computational domain is triply-periodic, with overall dimensions ($L_x, L_y, L_z$). Our choice for $L_x$, $L_y$ and $L_z$ is essentially dictated by computational constraints.  Since in stars the scale of the shear is expected to be much larger than the scale of individual fingers, we must respect that hierarchy here as well. As shown by \citet{Traxler2011b} the vertical scale of fingers is close to their horizontal scale $2\pi / \hat l_f$ where $\hat l_f$ is the wavenumber of the fastest-growing fingering mode (in the absence of shear). Since $\hat l_f \simeq 0.5$ in our selected units across almost all parameter space, we have to satisfy the condition $L_z \gg  2\pi / \hat l_f \sim 10$. For all the simulations shown below we have picked $L_z = 200$. When the forcing is strong (low Ri) and the density ratio is relatively high, the fingers are strongly stretched horizontally. In order to avoid being overly constrained by the domain size, most of our simulations are therefore run in long domains with $L_x = 500$ except where specifically mentioned. With such large values of $L_x$ and $L_z$, we have then been forced to use a relatively narrow domain in the remaining direction, choosing $L_y  = 25$. As discussed by \citet{GaraudBrummell2015}, the choice of $L_y$ in the fingering regime does not affect the results much as long as $L_y$ is larger than about 2 finger widths (hence our choice of $L_y$). Tables \ref{tab:data1} and \ref{tab:data2} summarize all the runs we have performed. 

Except when specifically mentioned, all the simulations were run until a statistically stationary state was reached, either starting from the laminar state with small added perturbations in the compositional field, or starting from the end-state of another simulation at nearby parameters (e.g. gradually increasing or decreasing either the Richardson number or the density ratio, as appropriate). We have found that the initial conditions used have no influence on the nature of the statistically stationary state reached by the simulation. Owing to the high computational costs of low Prandtl number / low diffusivity ratio simulations (where the viscous and compositional boundary layers at the edges of the fingers are very thin, and yet must be fully resolved), most of the simulations were run with ${\rm Pr} = \tau  = 0.3$. While this is far from stellar values, these parameters are smaller than one, as they would be in stellar interiors. A few additional runs at ${\rm Pr} = \tau  = 0.03$ are also presented to verify the adequacy of any model or scaling law we may derive from analyzing the ${\rm Pr} = \tau  = 0.3$ data. The numerical resolution used for each run (written in terms of equivalent grid points used) varies depending on the parameters selected, and is reported in each case below. We checked that for all the runs shown, the simulations are fully resolved.

\subsection{Results for ${\rm Pr} = \tau = 0.3$.}

We focus for now on a first set of simulations that have ${\rm Pr} = \tau  = 0.3$. In all cases the domain size is selected to have $L_x  = 500$, $L_y =  25$ and $L_z = 200$, and the simulations have been run until a statistically stationary state is reached. For cases where $R_0 = 1.5$, $1.75$ and $2$, the resolutions used is $(960 \times 48 \times 192)$ equivalent grid points, while for cases where $R_0 = 2.25$ and 2.5, the resolution used is $(480 \times 24 \times 96)$ equivalent grid points.  A summary of all the runs performed in this manner is presented in Table \ref{tab:data1}.

\begin{table}[]
        \caption{\small Summary of the parameters and main results for simulations with ${\rm Pr} = \tau = 0.3$. All measurements are taken as time averages once the system has reached a statistically stationary state. For each pair of parameters $(R_0, {\rm Ri})$, the third and fourth columns show the time-average of $\hat u_{\rm rms}(t)$ and $\hat w_{\rm rms}(t)$ (see equation \ref{eq:urmswrms}), and the errors represent a standard deviation around the mean. The fifth column represents the time-average of the maximum value of $\overline{\hat u}$, see text for detail, and the error represents a standard deviation around the mean. The sixth column represents the amplitude of the shear $\hat S_m$, see text for detail; the errorbar is the error on the fitting parameter. The seventh column represents the volume average of $|\nabla \hat C|^2$, and the error is the standard deviation about the mean.  Finally, the eighth column represents the turbulent viscosity estimated by fitting the relationship between $\overline{\hat u \hat w}$ and $\hat S$, see text for detail. If the relationship is linear, the errorbar is the error on the fitting parameter. If the relationship is nonlinear, the fit of the mean profile is reported first, and the bracket contain the minimum and maximum values of $\hat \nu_{\rm turb}$ measured in the center and wings of the profile (see text for detail). }
        \label{table1}
        \centering{
        \vspace{0.3cm}
        {\small 
\begin{tabular}{cccccccc}
                \tableline
                 $R_0$ & ${\rm Ri}$ &  $\hat u_{\rm rms}$        &       $\hat w_{\rm rms}$     &     max$\left(\overline{\hat u}\right)$  &   $\hat S_m$ &         $ \langle |\nabla \hat C|^2 \rangle $      & $\hat \nu_{\rm turb}$           \\
                        \tableline
  1.5 & 0.1 & 2.55 $\pm$ 0.01 & 1.36  $\pm$ 0.04 & 3.4  $\pm$   0.05 & 0.107 $\pm$  0.02 & 13.32  $\pm$ 0.24 & 2.46 $\pm$ 0.01 \\ 
 1.5 & 0.3 & 1.46 $\pm$ 0.006 & 1.37  $\pm$ 0.01 & 1.8  $\pm$   0.04 & 0.056 $\pm$  0.02 & 14.4  $\pm$ 0.29 & 2.72 $\pm$ 0.01 \\ 
 1.5 & 3 & 0.81 $\pm$ 0.01 & 1.39  $\pm$ 0.01 & 0.55  $\pm$   0.04 & 0.017 $\pm$  0.02 & 14.8  $\pm$ 0.28 & 2.8 $\pm$ 0.02 \\ 
 1.5 & 30 & 0.72 $\pm$ 0.007 & 1.39  $\pm$ 0.01 & 0.18  $\pm$   0.04 & 0.006 $\pm$  0.02 & 14.8  $\pm$ 0.30 & 2.23 $\pm$ 0.08 \\ 
 \\
 1.75 & 0.3 & 3.00 $\pm$ 0.001 &  0.809  $\pm$ 0.006 & 4.17  $\pm$   0.06 & 0.13 $\pm$  0.01 & 4.43  $\pm$ 0.08 & 1.18 $\pm$ 0.01 \\ 
 1.75 & 3 & 0.912 $\pm$ 0.005 & 0.983  $\pm$ 0.007 & 1.10 $\pm$   0.03 & 0.035 $\pm$  0.01 & 6.29  $\pm$ 0.11 & 1.49 $\pm$ 0.02 \\ 
 1.75 & 30 & 0.525 $\pm$ 0.004 & 0.99  $\pm$ 0.008 & 0.33  $\pm$   0.03 & 0.01 $\pm$  0.01 & 6.41  $\pm$ 0.11 & 1.48 $\pm$ 0.02 \\ 
 1.75 & 100 & 0.489 $\pm$ 0.004 & 0.99  $\pm$ 0.007 & 0.24  $\pm$   0.04 & 0.007 $\pm$  0.01 & 6.37  $\pm$ 0.18 & 1.52 $\pm$ 0.02 \\ 
\\
 2 & 0.1 & 11.46 $\pm$ 0.08 & 0.27  $\pm$ 0.01 & 16.3  $\pm$   0.1 & 0.51 $\pm$  0.03 & 0.37  $\pm$ 0.01 & 0.40 $\pm$ 0.01 \\ 
 2 & 0.3 & 5.81 $\pm$ 0.04 & 0.30  $\pm$ 0.005 & 8.2  $\pm$   0.1 & 0.25 $\pm$  0.02 & 0.60  $\pm$ 0.02 & 0.54 [0.30,0.68] \\ 
 2 & 1 & 2.80 $\pm$ 0.01 & 0.505  $\pm$ 0.004 & 3.9  $\pm$   0.1 & 0.12 $\pm$  0.01 & 1.61  $\pm$ 0.02 & 0.65 [0.42,0.92] \\ 
 2 & 3 & 1.74 $\pm$ 0.006 & 0.64  $\pm$ 0.004 & 2.42  $\pm$   0.04 & 0.076 $\pm$  0.007 & 2.41  $\pm$ 0.03 & 0.58 $\pm$ 0.01 \\ 
 2 & 10 & 0.95 $\pm$ 0.01 & 0.70  $\pm$ 0.005 & 1.26  $\pm$   0.03 & 0.04 $\pm$  0.006 & 2.86  $\pm$ 0.04 & 0.67 $\pm$ 0.01 \\ 
 2 & 30 & 0.59 $\pm$ 0.007 & 0.72 $\pm$ 0.006 & 0.7  $\pm$   0.02 & 0.02 $\pm$  0.007 & 2.96  $\pm$ 0.05 & 0.68 $\pm$ 0.01 \\ 
\\
 2.25 & 0.1 & 16.03 $\pm$ 0.03 & 0.18  $\pm$ 0.01 & 22.6  $\pm$   0.17 & 0.71 $\pm$  0.05 & 0.16  $\pm$ 0.01 & 0.24 $\pm$ 0.01 \\ 
 2.25 & 0.3 & 9.04 $\pm$ 0.01 & 0.17  $\pm$ 0.004 & 12.8  $\pm$   0.08 & 0.40 $\pm$  0.03 & 0.18  $\pm$ 0.008 & 0.27 [0.21,0.28] \\ 
 2.25 & 3 & 2.53 $\pm$ 0.006 & 0.275  $\pm$ 0.003 & 3.6  $\pm$   0.08 & 0.11 $\pm$  0.01 & 0.47  $\pm$ 0.01 & 0.33 [0.12,0.63] \\ 
 2.25 & 30 & 1.18 $\pm$ 0.004 & 0.43  $\pm$ 0.003 & 1.62  $\pm$   0.04 & 0.05 $\pm$  0.005 & 0.99  $\pm$ 0.01 & 0.13 [0.07,0.35] \\ 
 2.25 & 300 & 0.46 $\pm$ 0.006 & 0.49  $\pm$ 0.004 & 0.57  $\pm$   0.02 & 0.02 $\pm$  0.004 & 1.25  $\pm$ 0.02 & 0.09 $\pm$ 0.01 \\ 
\\
 2.5 & 0.1 & 19.0 $\pm$ 0.08 & 0.147  $\pm$ 0.015 & 26.8  $\pm$   0.26 & 0.84 $\pm$  0.07 & 0.09  $\pm$ 0.01 & 0.16 $\pm$ 0.01 \\ 
 2.5 & 0.3 & 11.37 $\pm$ 0.03 & 0.12  $\pm$ 0.005 & 16.0  $\pm$   0.12 & 0.50 $\pm$  0.03 & 0.09 $\pm$ 0.007 & 0.16 $\pm$ 0.01 \\ 
 2.5 & 3 & 3.38 $\pm$ 0.008 & 0.14  $\pm$ 0.003 & 4.77  $\pm$   0.04 & 0.15 $\pm$  0.007 & 0.125  $\pm$ 0.005 & 0.19 [0.06,0.23] \\ 
 2.5 & 30 & 1.30 $\pm$ 0.002 & 0.23  $\pm$ 0.003 & 1.83  $\pm$   0.05 & 0.06 $\pm$  0.005 & 0.28  $\pm$ 0.006 & 0.1 [-0.05,0.36] \\ 
 2.5 & 100 & 0.96 $\pm$ 0.001 & 0.27  $\pm$ 0.003 & 1.34  $\pm$   0.04 & 0.04 $\pm$  0.004 & 0.38  $\pm$ 0.007 & 0.03 [-0.09,0.26] \\ 
 2.5 & 1000 & 0.57 $\pm$ 0.001 & 0.32  $\pm$ 0.003 & 0.78  $\pm$   0.02 & 0.025 $\pm$  0.003 & 0.51  $\pm$ 0.009 & -0.12 [-0.17,0] \\ 
 2.5 & 10000 & 0.21 $\pm$ 0.001 & 0.34  $\pm$ 0.003 & 0.21  $\pm$   0.02 & 0.0065 $\pm$  0.003 & 0.57  $\pm$ 0.01 & -0.18 $\pm$ 0.01 \\ 
\tableline
\end{tabular}
}
}
\label{tab:data1}
\end{table}

\begin{figure}[h]
\epsscale{0.7}
\plotone{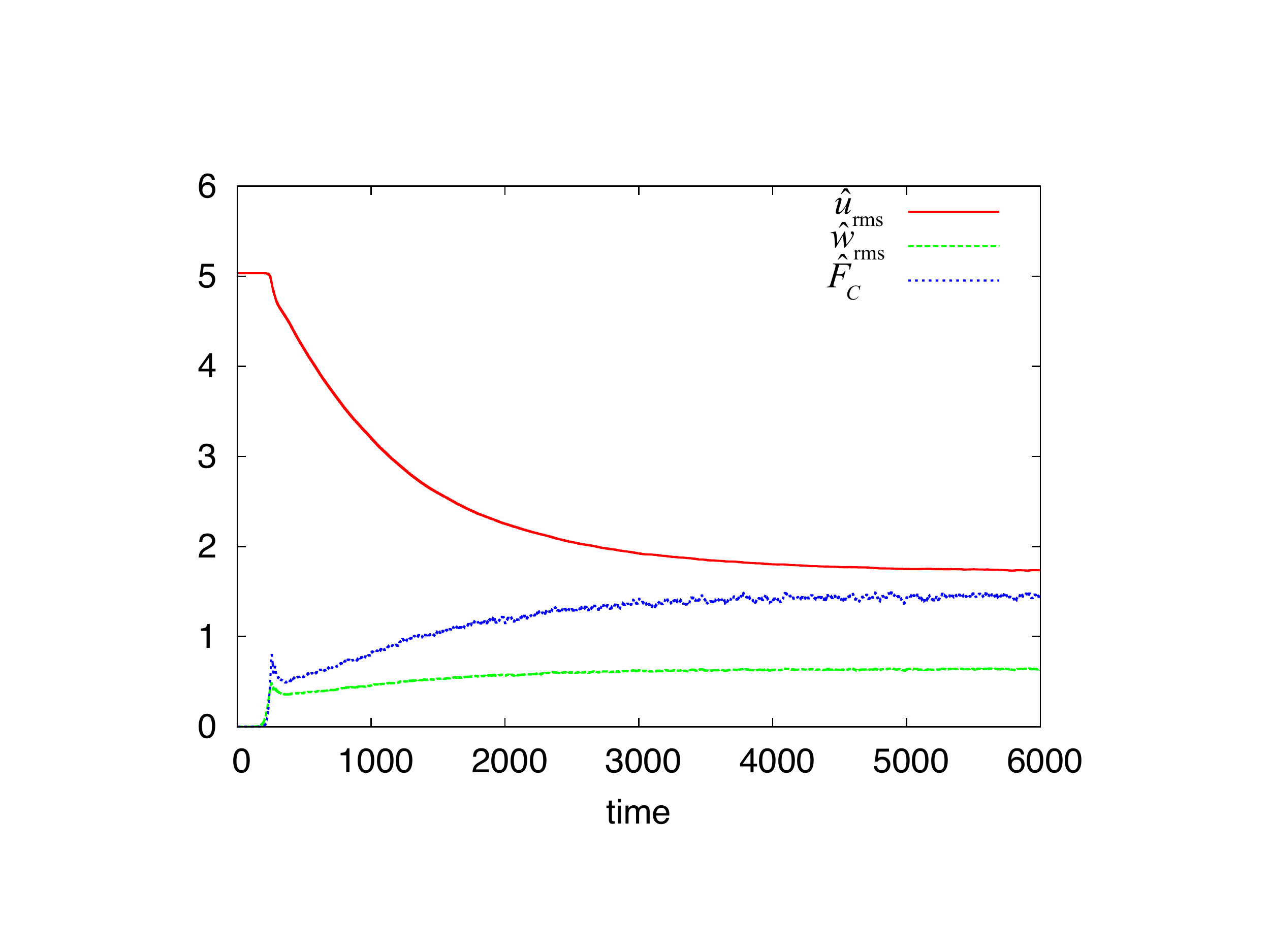}
\caption{Temporal evolution of $\hat u_{\rm rms}$, $\hat w_{\rm rms}$ and $\hat F_C$ (see text for definitions) in a simulation with ${\rm Pr} = \tau  = 0.3$, $R_0 = 2$ and ${\rm Ri} = 3$.}
\label{fig:timevol}
\end{figure}

Figure \ref{fig:timevol} shows the temporal evolution of a typical simulation in the shear-stable case, that was started from laminar initial conditions plus small-amplitude perturbations, and has a density ratio $R_0 = 2$ (which is in the middle of the fingering-unstable range for $R_0$ at $\tau  =0.3$) and ${\rm Ri} = 3$. Shown are the rms values of the streamwise and vertical velocities 
\begin{equation}
\hat u_{\rm rms}(t) = \langle \hat u^2 \rangle ^{1/2} \mbox{  and  } \hat w_{\rm rms}(t) = \langle \hat w^2 \rangle ^{1/2} ,
\label{eq:urmswrms}
\end{equation}
where $ \langle \cdot \rangle$ denotes a volume average over the computational domain. Also shown is a proxy for the absolute value of the compositional flux, which filters out rapid time variations due to internal gravity waves \citep[c.f.][]{Malkus1954,Woodal13} when the system is in a statistically stationary state: 
\begin{equation}
\hat F_C(t) = \left| \langle \hat w \hat C \rangle \right|  \simeq \tau R_0 \langle | \nabla \hat C |^2 \rangle .
\label{eq:FCdef}
\end{equation}
Note that $\langle \hat w \hat C \rangle$ in fingering convection is always negative, and so $\langle \hat w \hat C \rangle  = - \hat F_C$.

We see two phases of evolution in Figure \ref{fig:timevol}. The early time behavior is dominated by the development of the fingering instability on the background shear flow (at this value of ${\rm Ri}$, the shear instability is not active). We see that $\hat u_{\rm rms}(t)$ is initially steady while $\hat w_{\rm rms}(t)$ grows exponentially, until the perturbations gain sufficient amplitude to affect the mean flow (around $t = 30$). This marks the onset of the second phase, in which the mean streamwise flow and the fingering field interact nonlinearly with one another on a slower timescale. When this happens, the streamwise velocity begins to decrease, as a
result of the small-scale fingering which acts as an effective turbulent viscosity (more on this
below). Because the shear impedes mixing by fingering convection, proxies for the turbulence (namely $\hat w_{\rm rms}$ and $\hat F_C$) slowly grow with time as the shear decreases. This contest between the background shear and fingering convection eventually leads to a statistically stationary state.

\begin{figure}[h!]
\epsscale{0.9}
\plotone{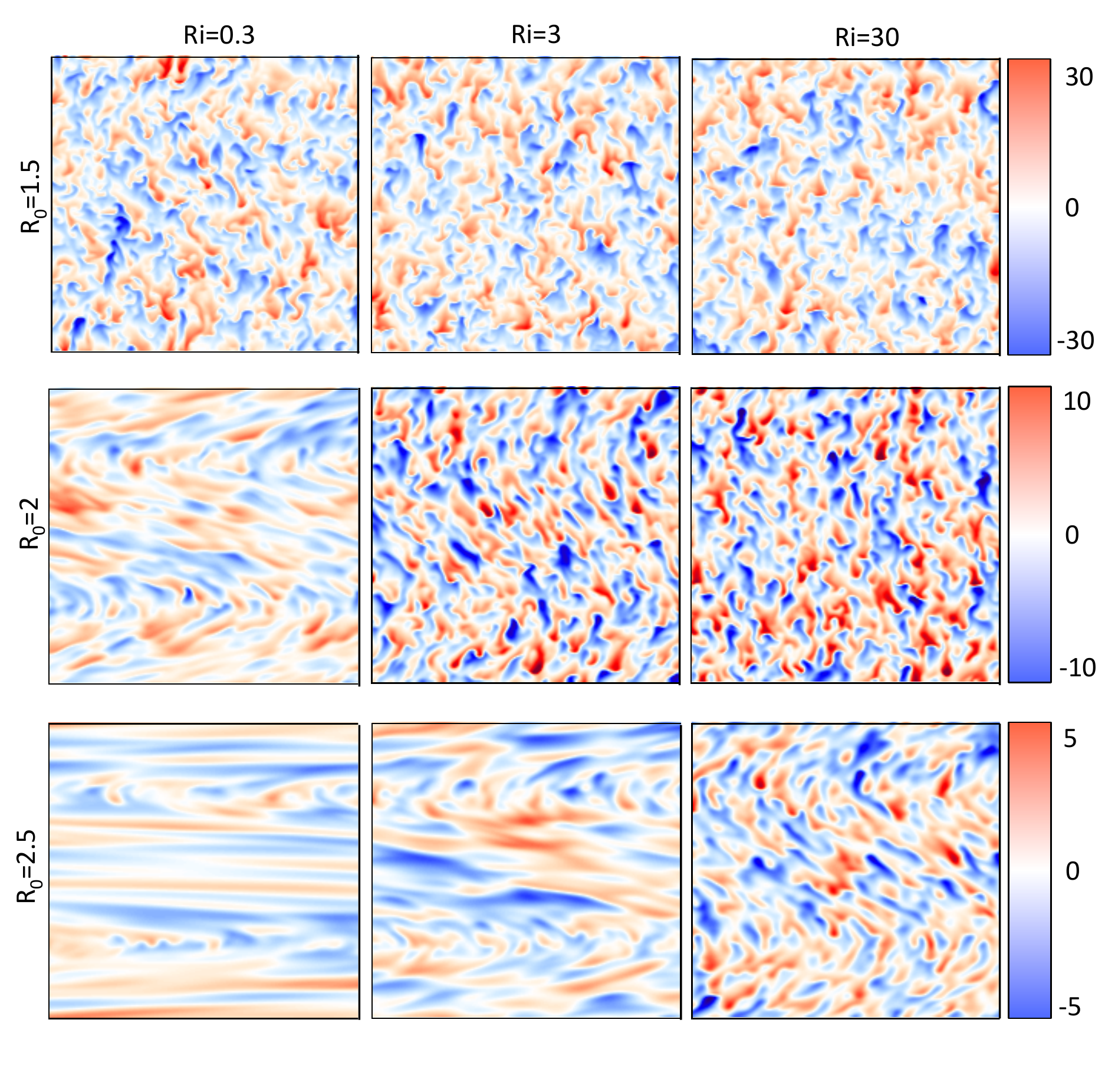}
\caption{Snapshots of the compositional perturbation $\hat C(x,0,z)$ in 9 different simulations, for varying $R_0$ and ${\rm Ri}$, taken once the system has achieved a statistically stationary state. Shown is the $y = 0$ plane. Note that each panel only shows part of the domain in the streamwise direction, for $x \in [0,200]$, instead of the whole domain.}
\label{fig:snaps}
\end{figure}

Figure \ref{fig:snaps} shows snapshots of the compositional perturbations in the statistically stationary state for three different values of the density ratio spanning the fingering instability range (which, at the selected values of Pr and $\tau$ is $R_0 \in [1,10/3]$), and three different values of the Richardson number. When the density ratio is close to the onset of convection ($R_0 = 1.5)$, the intrinsic growth rate of the fingering instability is large, and the shear does not affect the latter much even for small values of ${\rm Ri}$. By contrast, for a large density ratio closer to marginal stability for the fingering instability ($R_0 = 2.5)$, the shear clearly tilts the much more slowly growing fingers even for large Richardson number.

\begin{figure}[h!]
\epsscale{0.7}
\plotone{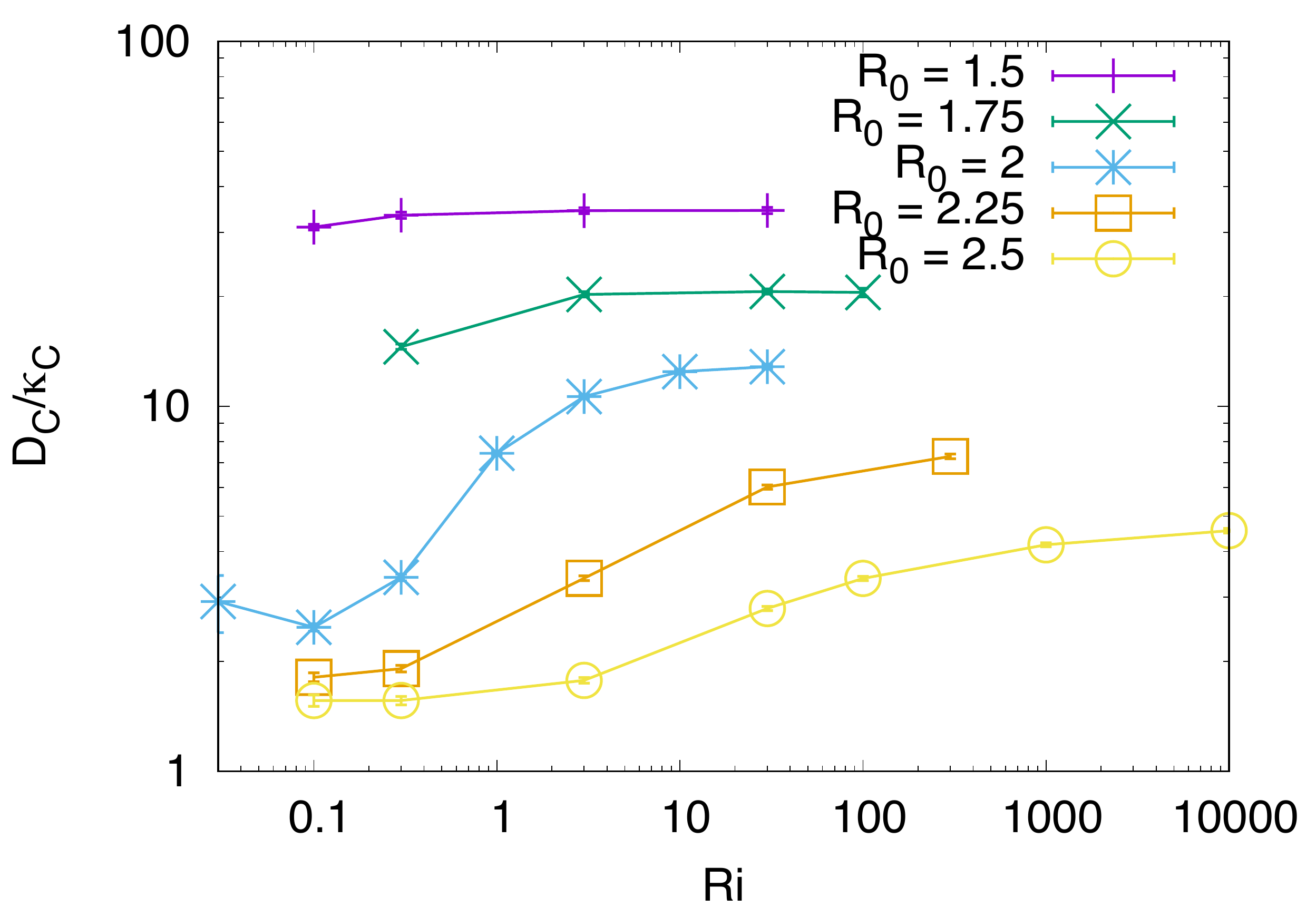}
\caption{Variation of the quantity $D_C / \kappa_C$ with density ratio $R_0$ and laminar Richardson number ${\rm Ri}$ for the simulations presented in Table \ref{tab:data1}. Error bars are shown for each data point, but they are typically smaller than the size of the symbol. For sufficiently large ${\rm Ri}$, $D_C / \kappa_C$ tends to the corresponding unsheared value.}
\label{fig:NucvsRi}
\end{figure}

The effect of the shear on vertical compositional mixing by fingering convection can easily be quantified by measuring the ratio
\begin{equation}
\frac{D_C}{\kappa_C} = \frac{\kappa_C + \kappa_{C,{\rm turb}}}{\kappa_C} = \frac{ - \kappa_C C_{0z} + \langle wC \rangle }{- \kappa_C C_{0z}} = 1 - \frac{R_0}{\tau} \langle \hat w \hat C \rangle = 1 + \frac{R_0 }{\tau} \hat F_C ,
\label{eq:39}
\end{equation}
where $\hat F_C$ is the compositional flux defined in equation (\ref{eq:FCdef}). This defines $D_C$ as the effective compositional diffusivity, which is the sum of the microscopic diffusivity $\kappa_C$ and the turbulent diffusivity $\kappa_{C,{\rm turb}} = - \langle wC \rangle / C_{0z}$. The ratio $D_C/\kappa_C$ is often called the compositional Nusselt number. Note that the first three expressions in (\ref{eq:39}) contain only dimensional quantities, while the last two contain non-dimensional quantities.  Figure \ref{fig:NucvsRi} presents the time average of $D_C/\kappa_C$, measured once the system has reached a statistically stationary state. We see that decreasing ${\rm Ri}$ (i.e. increasing the shear) always acts to reduce $D_C/\kappa_C$ (except at very low values of the Richardson number where shear instabilities are finally excited, not shown here). However, we also see that the effect of the shear on the fingers is almost negligible at low density ratios, but becomes very significant at higher density ratios, where the flux is reduced even at large values of ${\rm Ri}$. This is consistent with the visual inspection of the simulation snapshots shown in Figure \ref{fig:snaps}. A simple model for the reduction of the compositional flux in the presence of shear is presented in Section \ref{sec:simplemodel}.

\begin{figure}
\epsscale{0.9}
\plotone{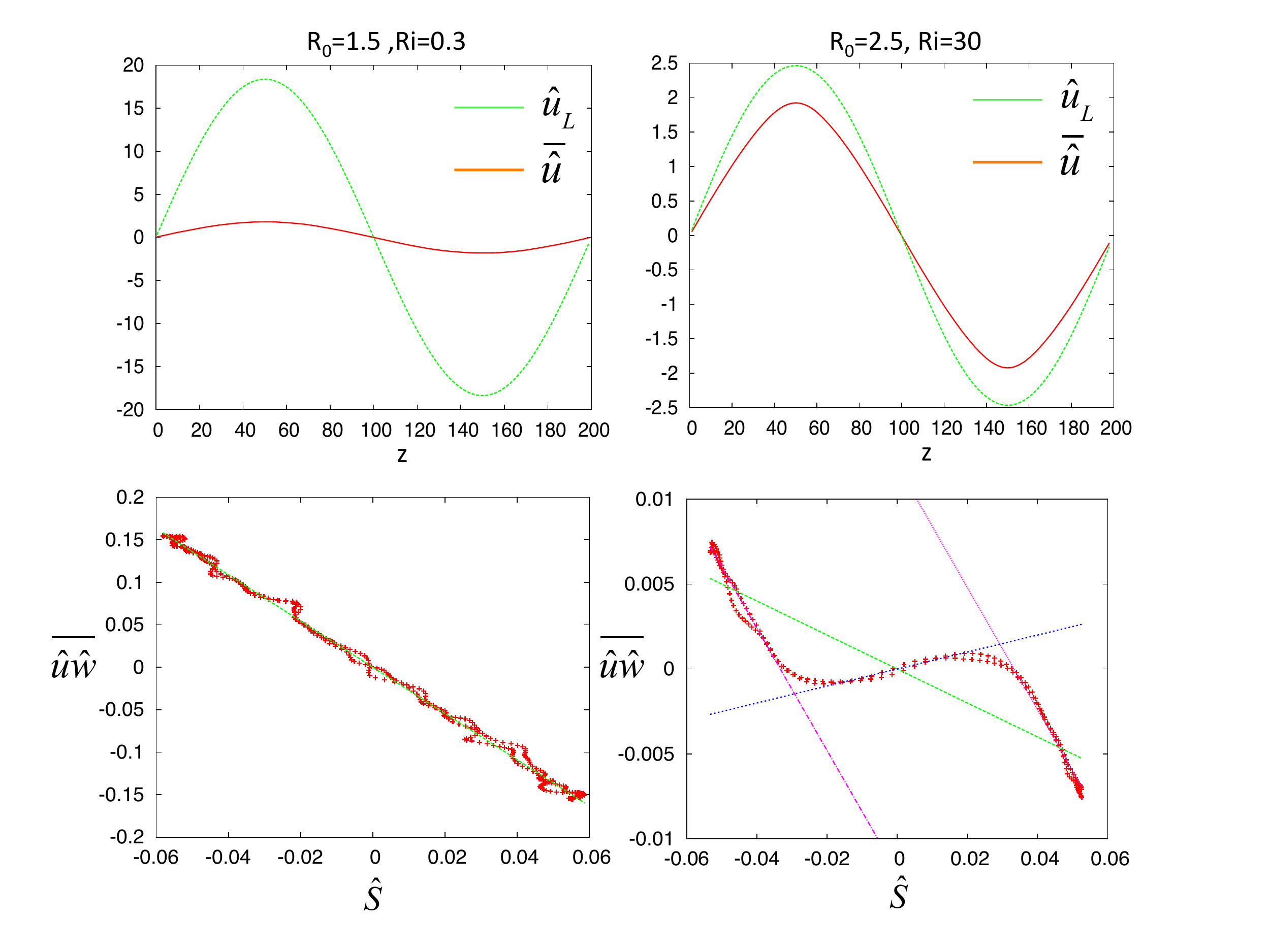}
\caption{Top row:  Comparison between the laminar flow $\hat u_L \sin(\hat k_sz)$ (green line) and the horizontally averaged, time averaged streamwise velocity $\overline{\hat u}(z)$ (red line) for two of the simulations presented in Figure \ref{fig:snaps}. The left column shows data for $R_0 = 1.5$, ${\rm Ri} = 0.3$ while the right column shows data for  $R_0 = 2.5$, ${\rm Ri} = 30$. Bottom row: Parametric plots of the Reynolds stress $\overline{\hat u \hat w}(z)$ against the mean shear $\hat S(z)$ (see text for exact definition), in the same two simulations. In both plots, the green line shows a linear fit to the full dataset. On the right, also shown are the linear fit to the same relationship but for small shear only (blue line) and for large shear only (pink line). Anti-diffusive behavior is observed for small shearing rates in this simulation.}
\label{fig:umean}
\end{figure}

Turbulent fingering convection does not only transport chemical elements, but also transports momentum. This could be seen for instance in the response of the mean flow to the development of the fingering instability in Figure \ref{fig:timevol}. Figure \ref{fig:umean} (top row) shows the horizontally averaged and time-averaged streamwise velocity $\overline{ \hat u}$ as a function of $z$, for two of the simulations that were shown in Figure \ref{fig:snaps}, namely one with $R_0 = 1.5$ and ${\rm Ri} = 0.3$, where the fingering is strong, and one with $R_0 = 2.5$ and ${\rm Ri} = 30$, where the fingering is weak. As usual, the time average is taken once the simulation is in a statistically stationary state. Also shown is the corresponding laminar mean  flow $\hat u_L \sin(\hat k_s z)$ for the same simulation. In both cases, we see that the turbulent mean flow has a similar sinusoidal shape to that of the laminar flow (though perhaps slightly more triangular in the $R_0 = 2.5$ and ${\rm Ri} = 30$ case), but has a different amplitude. This difference can be used to estimate the turbulent viscosity of the system. In a statistically stationary state, the horizontal average of the momentum equation is
\begin{equation}
\frac{\partial}{\partial z} \overline{\hat u \hat w} = {\rm Pr} \frac{\partial^2 \overline{\hat u}}{\partial z^2} + \hat F_0 \sin(\hat k_s z). 
\end{equation}
Assuming, as it is common to do so, that the Reynolds stress can be written as
\begin{equation}
 \overline{\hat u \hat w} = - \hat \nu_{\rm turb} \frac{\partial \overline{\hat u}}{\partial z} ,
 \label{eq:assumeRey}
\end{equation}
(where $\hat \nu_{\rm turb}$ is a non-dimensional turbulent viscosity) then the momentum balance reduces to 
\begin{equation}
( {\rm Pr} + \hat \nu_{\rm turb})  \frac{\partial^2 \overline{\hat u}}{\partial z^2} = - \hat F_0 \sin(\hat k_s z) ,
\end{equation}
whose solution is 
\begin{equation}
 \overline{\hat u}(z) = \frac{ \hat F_0}{({\rm Pr} +\hat \nu_{\rm turb}) \hat k_s^2 }  \sin(\hat k_s z) .
 \end{equation}
Comparing this expression with the definition of the laminar flow, we should therefore have 
\begin{equation}
\frac{ \hat u_L(z)}{ \overline{\hat u}(z)} = \frac{{\rm Pr} + \hat \nu_{\rm turb}}{{\rm Pr} } ,
\label{eq:nuturb1}
\end{equation}
as long as the assumption made in (\ref{eq:assumeRey}) applies. Equation (\ref{eq:nuturb1}) can be used to estimate $\hat \nu_{\rm turb}$ from the data (see below).

However, a direct inspection of the relationship between the mean shear and the Reynolds stress $\overline{\hat u \hat w} $ reveals that (\ref{eq:assumeRey}) does not always apply.
The bottom panels of Figure \ref{fig:umean} show the time average of the Reynolds stress $\overline{\hat u \hat w}(z)$ against $\hat S(z) \equiv \partial \overline{\hat u}/\partial z$, as a parametric plot with $z$ being the parameter.  According to (\ref{eq:assumeRey}) these two quantities should be linearly related to one another, and we see that this is indeed the case for the simulation with $R_0 = 1.5$ and ${\rm Ri} = 0.3$. However the $R_0 = 2.5$ and ${\rm Ri} = 30$ simulation behaves very differently, and reveals a strongly nonlinear relationship between $\overline{\hat u \hat w}$ and $\hat S$. This nonlinear behavior is in fact common for all simulations with relatively low shear and high density ratios. In extreme cases, such as the $R_0 = 2.5$, ${\rm Ri} = 30$ case shown here, fingering convection can even behave anti-diffusively, i.e. with $\overline{\hat u \hat w}$ locally increasing with $\hat S$. This recovers a relatively well-known result first obtained by \citet{Holyer1984} and discussed in various other publications since \citep[e.g.][]{PaparellaSpiegel1999,Xieal2019}.  Anti-diffusive tendencies only happen for relatively high ${\rm Ri}$ cases, however. For higher shearing rates, high density ratio simulations have a $\overline{\hat u \hat w}$ vs. $\hat S$ relationship that is once again both linear and decreasing.

These finding suggest two ways of estimating the turbulent diffusivity: one using the stress-strain relationship (\ref{eq:assumeRey}), and one using the mean flow, from (\ref{eq:nuturb1}). 
In the latter case, we measure the amplitude of $ \overline{\hat u}(z)$ by fitting the profile to a sinusoidal function. We then solve (\ref{eq:nuturb1}) for $\hat \nu_{\rm turb}$ using that measured amplitude. 
In the former case, we compute the time average of the Reynolds stress profile $\overline{\hat u \hat w}(z)$ as well as the vertical derivative of the mean flow $\overline{\hat u}$ to compute $\hat S(z)$, as was done in Figure \ref{fig:umean} (bottom panels). We then fit a linear function of the kind 
\begin{equation}
 \overline{\hat u \hat w} = - \hat \nu_{\rm turb}\hat S + b , 
 \label{eq:linearfit}
 \end{equation}
to this data. For cases where the $\overline{\hat u \hat w}$ vs. $\hat S$ relationship is linear, we set $b = 0$. For cases where the $\overline{\hat u \hat w}$ vs. $\hat S$ relationship is not linear, three fits are carried out. The first one, with $b = 0$, is a fit for the whole dataset, and reports a mean $\hat \nu_{\rm turb}$. Note that for a few simulations this measurement yields a negative value of $\hat \nu_{\rm turb}$. The second one, with $b = 0$, is taken for low values of $\hat S$ only, and reports a minimum value of $\hat \nu_{\rm turb}$, that is in some cases negative as well. The third fit is made with $b \neq 0$ to relate $|\overline{\hat u \hat w}|$ to $|\hat S|$ for large values of $|\hat S|$ only, which captures both regions of large positive and negative shear at the same time. This final fit estimates a maximum value of $\hat \nu_{\rm turb}$. Examples of the results of such fits are shown in Figure \ref{fig:umean} (bottom panels), for both the linear case and the nonlinear case. Table \ref{tab:data1} presents the results for each simulation.

\begin{figure}[h]
\epsscale{0.6}
\plotone{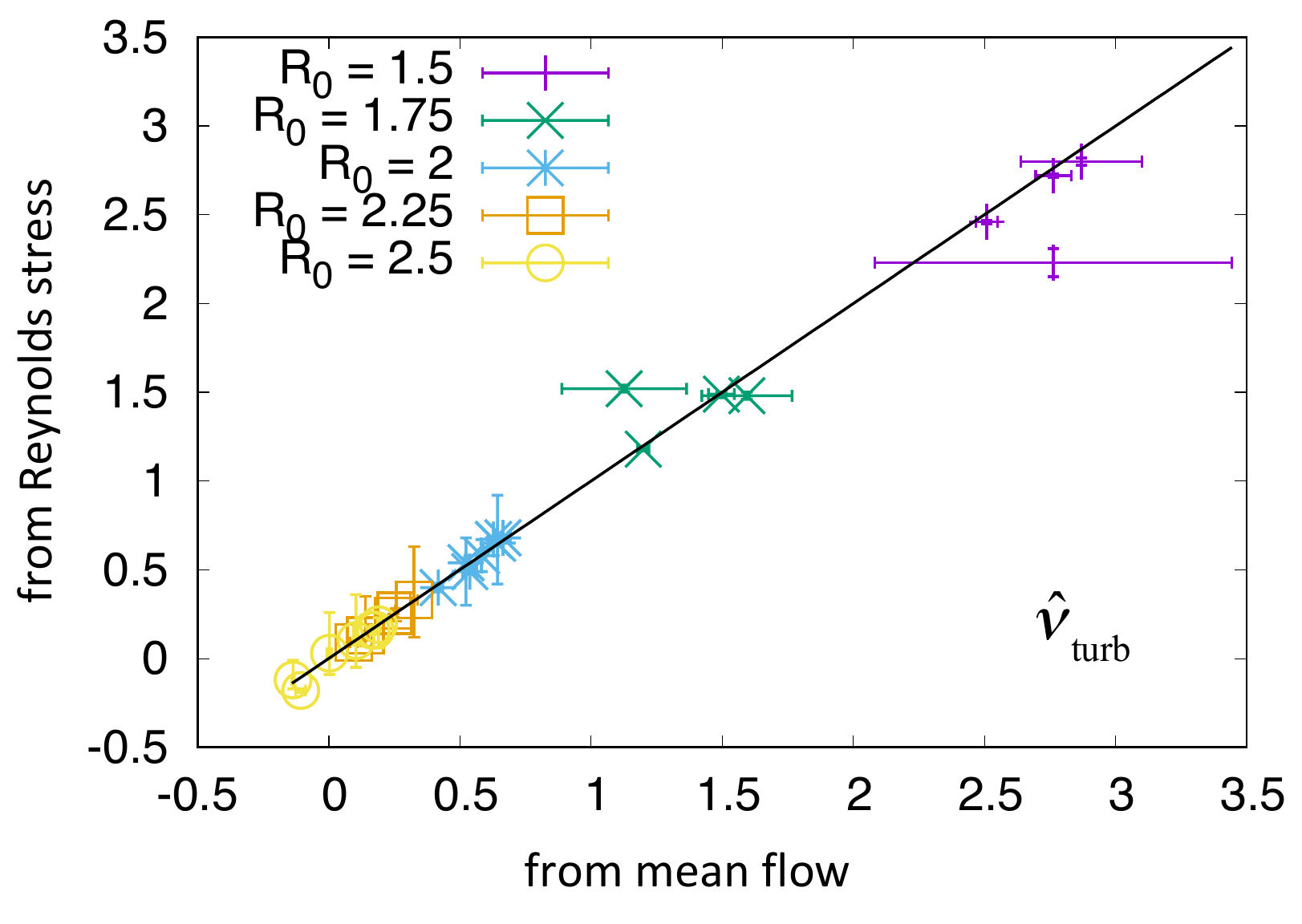}
\caption{Comparison between two methods of measuring $\hat \nu_{\rm turb}$ from the simulations presented in Table \ref{tab:data1}. The horizontal axis shows the results from the method using the mean flow using (\ref{eq:nuturb1}), and the vertical axis shows the results using the stress-strain relation using (\ref{eq:linearfit}). Error bars shown are computed as discussed in the main text.  }
\label{fig:Dnu}
\end{figure}

Figure \ref{fig:Dnu} compares the two methods of estimating the turbulent diffusivity to one another, with the mean-flow based method (using equation \ref{eq:nuturb1}) on the horizontal axis, and the stress-strain based method (using equation \ref{eq:linearfit}) on the vertical axis. The horizontal errorbar is computed from the measured error on the amplitude of $ \overline{\hat u}(z)$. The vertical errorbar is computed either from the error on the linear fit for cases where the relationship between $\overline{\hat u \hat w}$ and $\hat S$ is linear (typically that error is smaller than the symbol size), or ranges from the minimum to the maximum value of $\hat \nu_{\rm turb}$, with the symbol placed at the mean $\hat \nu_{\rm turb}$, for cases where the relationship between $\overline{\hat u \hat w}$ and $\hat S$ is nonlinear (see above). For the low density ratio simulations, $R_0 = 1.5$ and $R_0 = 1.75$, the $\overline{\hat u \hat w}$ vs. $\hat S$ relationship is linear for all values of ${\rm Ri}$, and the two methods of computing $\hat \nu_{\rm turb}$ are consistent except for two cases. These are the two simulations with largest Richardson numbers, which have very large horizontal errorbars because the horizontal mean flow is very weak. For these runs the mean-flow based method is inaccurate. For simulations at intermediate ($R_0 = 2$) and high ($R_0 = 2.25$ and $R_0 = 2.5$) density ratio, we begin to see examples of nonlinear relationship between $\overline{\hat u \hat w}$ and $\hat S$ especially at intermediate values of ${\rm Ri}$. Note that again for very large ${\rm Ri}$, the data is difficult to interpret as both $\overline{\hat u \hat w}$ and $\hat S$ are small compared with the fingering noise. Despite the nonlinearity of the stress-strain relationship, however, we see that the values of $\hat \nu_{\rm turb}$ computed using the mean flow method compares well with that computed by fitting a linear function to the full $\overline{\hat u \hat w}$ vs. $\hat S$ data (regardless of whether it is actually linear or not). This suggests that the details of the stress-strain relationship are less important than the mean behavior in setting the mean flow amplitude. 

Finally, as discussed above we see that for the largest density ratio simulations the mean value of $\hat \nu_{\rm turb}$ becomes slightly negative (i.e. the turbulence acts in an anti-diffusive way on average), and would on its own act to enhance the shear rather than suppress it. However, we also see that in all these anti-diffusive cases $| \hat \nu_{\rm turb}| < {\rm Pr} =  0.3$, so the system overall still behaves diffusively overall (i.e. with ${\rm Pr} + \hat \nu_{\rm turb} > 0$). In fact, so far we have found that  $| \hat \nu_{\rm turb}| \ll {\rm Pr}$ in all such runs, so the anti-diffusive effect is in practice negligible. Whether this would continue to be the case in stellar interiors remains to be determined, but is quite likely since the turbulent velocities are expected to scale as $\sqrt{{\rm Pr}/(R_0-1)}$ (in the pure fingering case at least) so the Reynolds stresses should scale as ${\rm Pr}/(R_0-1)$ \citep[see, e.g.][]{Brownal2013,SenguptaGaraud2018}. For large $R_0$ approaching marginal stability, which is the region of parameter space where anti-diffusive behavior may be expected \citep{Xieal2019}, the Reynolds stress is therefore expected to be very small. For this reason, we now set aside the mathematically interesting but in practice probably irrelevant scenario of anti-diffusive fingering convection.

\begin{figure}[h]
\epsscale{0.6}
\plotone{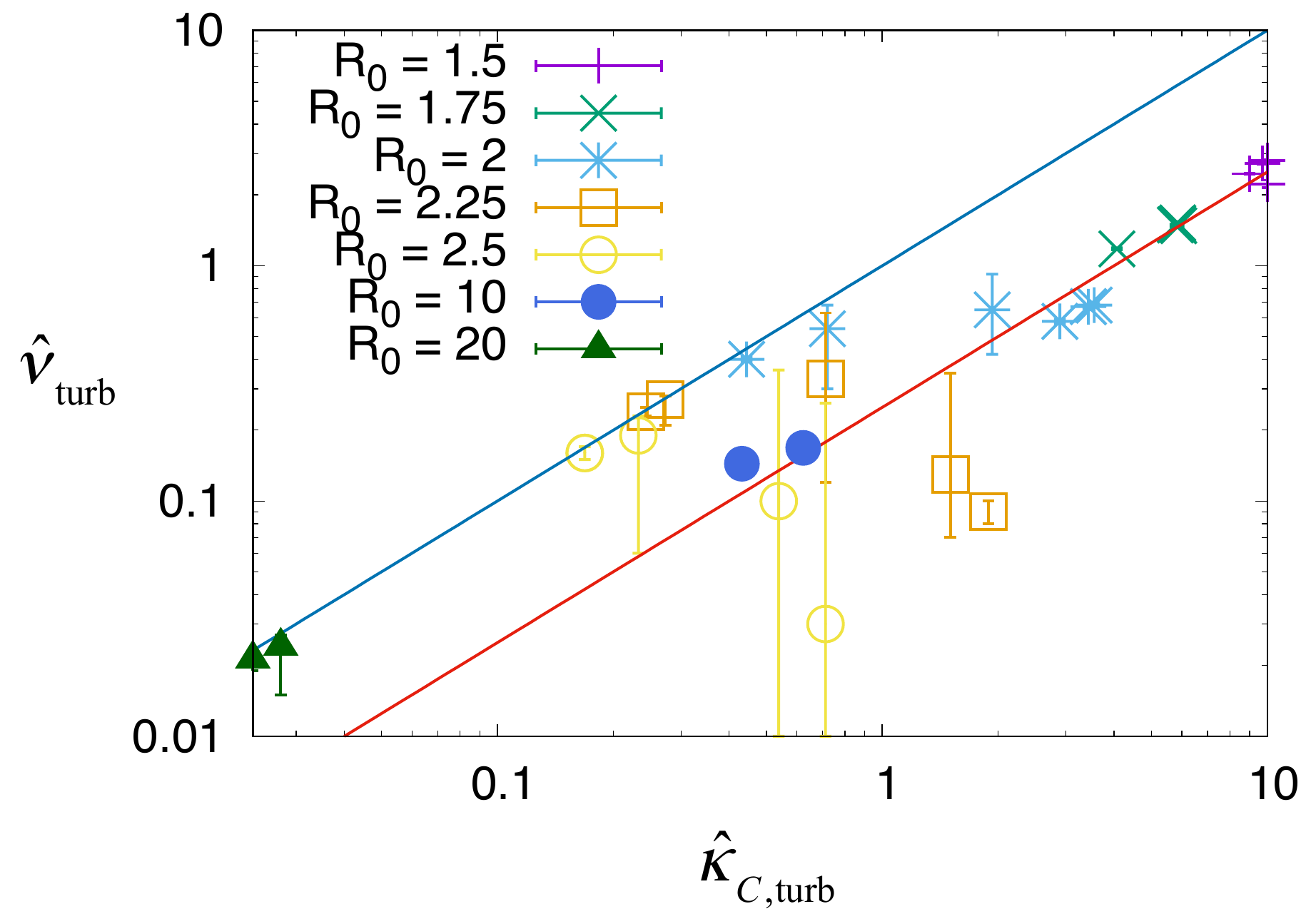}
\caption{Comparison between $\hat \nu_{\rm turb}$ and $\hat \kappa_{C,{\rm turb}}$ for a wide range of simulations, in statistically stationary state. Cases with $R_0 < 3$ are for ${\rm Pr} = \tau  = 0.3$, from the simulations described in Section \ref{sec:num} and Table \ref{tab:data1}. The $R_0 = 10$ and $R_0 = 20$ sets of simulations are for ${\rm Pr} = \tau = 0.03$, and are described in Section \ref{sec:simplemodel} and Table \ref{tab:data2}. For fixed values of $R_0$, increasing shear (lower Ri) reduces $\hat \kappa_{C,{\rm turb}}$. Vertical errorbars on $\hat \nu_{\rm turb}$ are computed as described in the main text. Horizontal errobars on $\hat \kappa_{C,{\rm turb}}$  are not shown as they would be smaller than the symbol size. The blue line shows the relationship $\hat \nu_{\rm turb} \simeq \hat \kappa_{C,{\rm turb}}$, and the red line shows $\hat \nu_{\rm turb} \simeq 0.25 \hat \kappa_{C,{\rm turb}}$.  }
\label{fig:DnuvsDC}
\end{figure}

A more practical result can be obtained by comparing the turbulent viscosity to the turbulent compositional diffusivity computed from the simulations, which is related to the flux $\hat F_C$ defined earlier in equation (\ref{eq:FCdef})
via 
\begin{equation}
\hat \kappa_{C,{\rm turb}} = \frac{- \langle wC \rangle}{\kappa_T C_{0z}} = R_0 \hat F_C.
\end{equation}
 Note that the second term in that expression is the ratio of two dimensional quantities, while the third is non-dimensional. Figure \ref{fig:DnuvsDC} shows the time-average of $\hat \kappa_{C,{\rm turb}}$ (during the statistically stationary state) against $\hat \nu_{\rm turb}$. We see that the data follows some interesting trends, and falls into two categories. Simulations that are clearly fingering-dominated, such as those with low density ratio ($R_0 = 1.5$ and $R_0 = 1.75$), or intermediate density ratio ($R_0 = 2$) but low shear, satisfy the relationship $\hat \nu_{\rm turb} \simeq 0.25 \hat \kappa_{C,{\rm turb}}$ (red line). By contrast, simulations that are clearly shear-dominated, such as those with high density ratio and high shear, satisfy the relationship $\hat \nu_{\rm turb} \simeq \hat \kappa_{C,{\rm turb}}$ (blue line). For some values of the density ratio (especially $R_0 = 2$), the data spans both limits, and continuously moves from one to the other as the Richardson number decreases (i.e. moving from right to left on the figure). The only data points that do not fall on either lines (or in between them) correspond to simulations with high density ratio and low shearing rates. In almost all cases, these also correspond to parameters where anti-diffusive behavior is observed, so $\hat \nu_{\rm turb}$ cannot remain proportional to $\hat \kappa_{C,{\rm turb}}$ in that limit\footnote{By construction, $\hat \kappa_{C,{\rm turb}}$ has to be positive, since $\langle \hat w \hat C \rangle$ has to be negative, so it cannot remain proportional to $\hat \nu_{\rm turb}$ when $\hat \nu_{\rm turb}$ changes sign.}. Since this behavior is unlikely to persist at stellar values of the Prandtl number, these data points a probably not significant for astrophysical purposes.

These results are important because they relate the momentum transport to the compositional transport. Hence if one can somehow be derived from observations (of, e.g. surface abundances or subsurface velocity profiles), the other can be indirectly inferred. 

\section{Model}
\label{sec:simplemodel}

We now propose a simple model to explain the trends observed in the previous section, which extends the work of \citet{Brownal2013} in the presence of a large-scale shear. For pedagogical purposes, we first briefly recall the salient properties of the original \citet{Brownal2013} model for mixing by fingering convection, and then add the shear. 

Given that the saturation of the fingering instability occurs as a result of parasitic shear instabilities that develop between up-flowing and down-flowing fingers, a key ingredient of the \citet{Brownal2013} model lies in assuming that this saturation occurs when the growth rate of the parasitic instabilities $\hat \sigma$ equals a universal constant times the growth rate of the original fingering instability $\hat \lambda_f$, i.e. when $ \hat \lambda_f = C_B \hat \sigma$. Furthermore, from dimensional analysis (or exact computations), it can be shown that $\hat \sigma = \eta \hat w_f \hat l_f$, where $\eta$ is a known constant (whose value is irrelevant). Hence the model predicts a simple relationship between the vertical velocity within the fingers $\hat w_f$, their growth rate $\hat \lambda_f$, and their wavenumber $\hat l_f$, namely 
\begin{equation}
 \hat \lambda_f = C_B \eta \hat w_f \hat l_f.
 \label{eq:parasitic}
\end{equation}
This idea was successfully verified by \citet{SenguptaGaraud2018}. Next, using the fact that 
\begin{equation}
\hat \lambda_f \hat C_f  + R_0^{-1} \hat w_f = - \tau \hat l_f^2 \hat C_f ,
\end{equation}
from equation (\ref{eq:nondimcomposition}) in the absence of shear, \citet{Brownal2013} estimated the amplitude of the compositional perturbation $\hat C_f$ to be 
\begin{equation}
\hat C_{f} = - \frac{R_0^{-1} \hat w_{f}  }{ \hat  \lambda_f   + \tau \hat l_f^2} .
\end{equation}
Finally, by dimensional analysis the compositional flux must be proportional to $\hat w_f \hat C_f$, which implies that 
\begin{equation}
\hat F_C  = | \langle \hat w \hat C \rangle | = K_B | \hat w_f \hat C_f |  =  K_B  \frac{R_0^{-1} \hat w^2_{f}  }{ \hat \lambda_f   + \tau \hat l_f^2}  = \frac{K_B}{(C_B \eta)^2} \frac{R_0^{-1} \hat \lambda^2_{f}  }{ \hat  \lambda_f \hat l_f^2   + \tau \hat l_f^4}, 
\label{eq:brownflux}
\end{equation}
where $K_B$ is, again, a universal constant. Note how all these universal constants now conveniently combine into a single one, that can be calibrated against the data. The model was successfully validated by \citet{Brownal2013} against DNSs of unsheared fingering convection with the constant $K_B/(C_B\eta)^2 \simeq 49$.  

\begin{figure}[h]
\epsscale{0.6}
\plotone{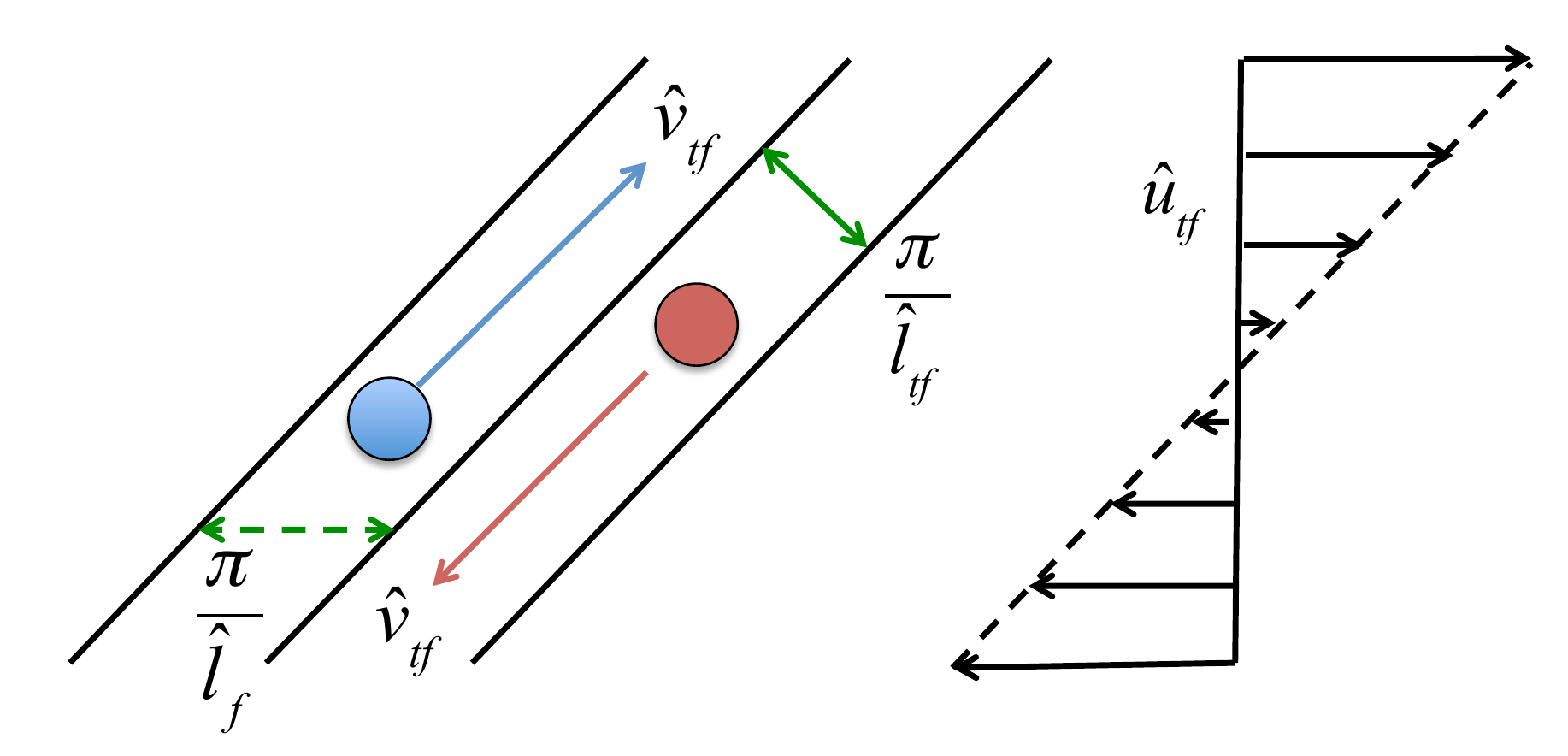}
\caption{Illustration of fingering convection in the presence of a uniform shear. The originally vertical finger, whose wavenumber is $\hat l_f$, is tilted by the shear, causing its width to decrease. The tilt also causes the addition of a horizontal velocity component, so the total velocity shear across two fingers would increase relative to the non-tilted one. Saturation by parasitic instabilities therefore occurs {\it earlier} in tilted fingers, reducing the vertical velocities and the turbulent flux.  }
\label{fig:tilt}
\end{figure}

In the presence of shear, the model must be modified to account for the fact that the fingers become tilted. This changes both their intrinsic velocity (which acquires a horizontal component) and their wavenumber (which decreases as a result of the tilt), see Figure \ref{fig:tilt}. To include this in the model, we begin with a thought experiment in which the fingers initially develop as they would without the shear (i.e. they grow at the same rate $\hat \lambda_f$ and have the same wavenumber $\hat l_f$ as the unsheared fingers), but are then 
subject to a homogeneous shear flow with constant shearing rate $\hat S$, whose effect is to tilt them away from the vertical. 

\begin{figure}[h]
\epsscale{0.6}
\plotone{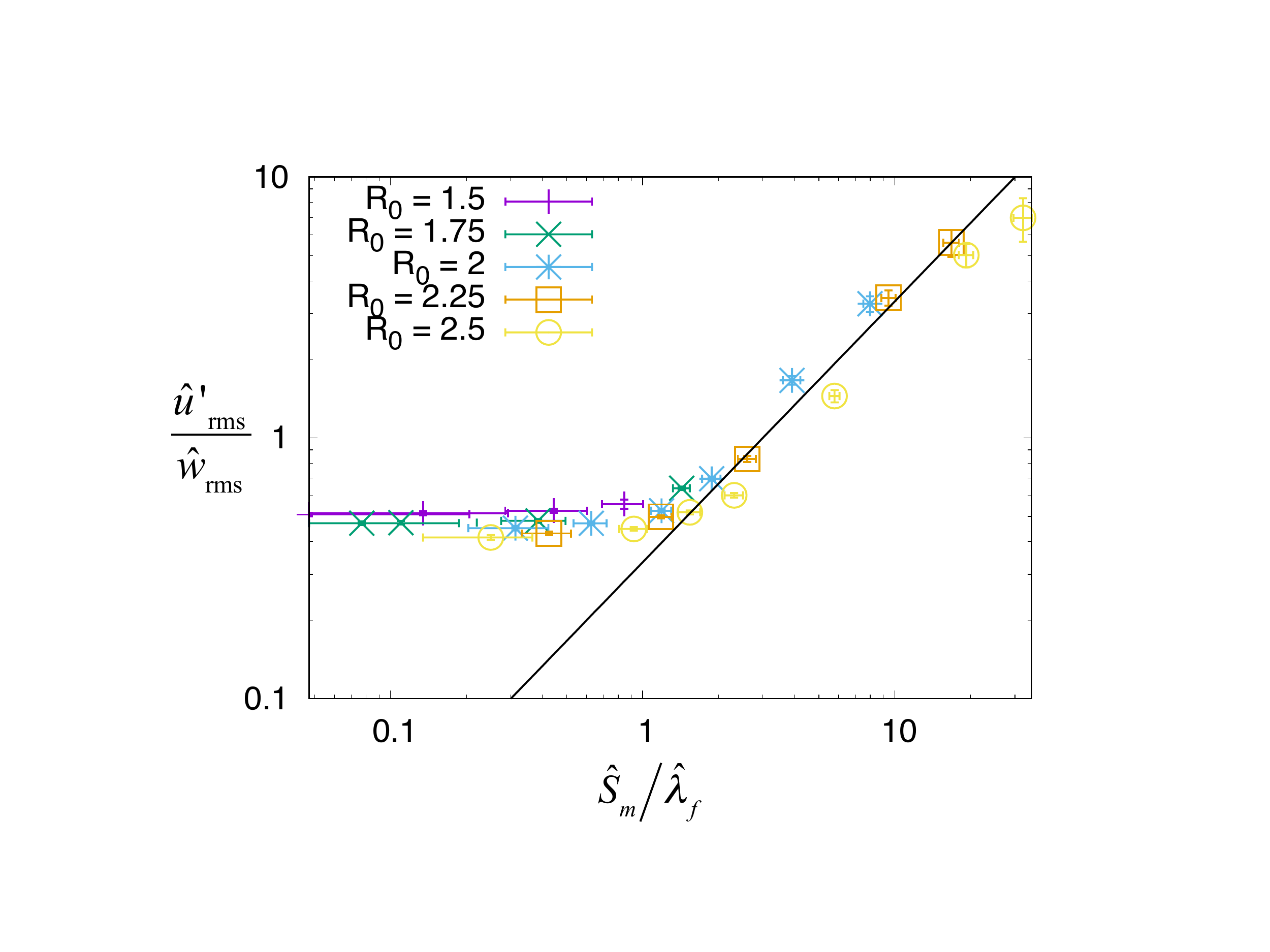}
\caption{Ratio of $\hat u'_{\rm rms}$ with $\hat w_{\rm rms}$, against $\hat S_m/\hat \lambda_f$. These quantities are defined in the text. The black line is the relationship $ \hat u'_{\rm rms}/\hat w_{\rm rms}= \hat S_m/ 3 \hat \lambda_f$. }
\label{fig:urmswrms}
\end{figure}

With this assumption, a fluid parcel that would be flowing purely vertically in the absence of shear acquires a horizontal velocity relative to the mean flow given by
\begin{equation}
\hat u_{tf} = \hat S dz = \hat S \hat w_{tf} dt,
\end{equation}
where $dz$ is the distance it has travelled, $\hat w_{tf}$ is its vertical velocity, and $dt$ is the travel time. The subscript ``$tf$" refers to the tilted finger (by contrast with the subscript ``$f$" which refers to the non-sheared vertical finger). We then assume that all parcels travel on average for a time $dt$ before losing coherence through nonlinear effects, and that this time is inversely proportional to the growth rate of the fingers $\hat \lambda_f$ (the reason for this assumption stems from the work of \citet{Brownal2013} described above). With this, we predict that 
\begin{equation}
\frac{\hat u_{tf}}{  \hat w_{tf} }  =  \chi \frac{\hat S}{ \hat \lambda_f}, 
\label{eq:ufwf}
\end{equation}
where $\chi$ is a universal constant of order unity. Note that up-flowing parcels thus acquire a positive horizontal velocity relative to the mean flow in this model, while downflowing parcels acquire a negative horizontal velocity relative to the mean flow. This relationship (and its underlying assumptions) can in fact be verified from the data. To do so, we compute an estimate of the typical vertical finger velocity $\hat w_{tf}$ in equation (\ref{eq:ufwf}) using the rms vertical velocity within the domain, namely $\hat w_{\rm rms}$ defined in equation (\ref{eq:urmswrms}). We also estimate the typical horizontal velocity of parcels $\hat u_{tf}$ relative to the background from 
\begin{equation}
\hat u'_{\rm rms} = \langle (\hat u - \overline{\hat u})^2 \rangle^{1/2}.
\end{equation}
Finally, the shear $\hat S$ is estimated by fitting a sinusoidal function to $\partial \bar u/ \partial z$, and reporting its amplitude (which corresponds to the maximum value of the shear in the flow) $\hat S_m$. 
Figure \ref{fig:urmswrms} shows $ \hat u'_{\rm rms} / \hat w_{\rm rms}$ as a function of $\hat S_m/\hat \lambda_f$ for all the simulations of Table \ref{tab:data1}, and we see that the data collapses onto a single universal curve, with two distinct regimes. For low shearing rates (relative to $\hat \lambda_f$), $\hat u'_{\rm rms}/\hat w_{\rm rms} \simeq 0.5$ is independent of the shear. This corresponds to the fingering-dominated limit, where any horizontal motion of the fluid is simply due to nonlinear saturation of the fingering instability rather than the tilting of the fingers. In that limit, $\hat u'_{\rm rms}$ is not a good predictor for $\hat u_f$, which is expected. On the other hand for larger shearing rates (when  $\hat S_m/ \hat \lambda_f$ is larger than one), then we see that $\hat u'_{\rm rms}/\hat w_{\rm rms} \simeq \hat S_m/ 3 \hat \lambda_f$, which confirms that $\hat u'_{\rm rms}$ is now indeed originating from the shear itself, and that equation (\ref{eq:ufwf}) applies with $\chi \simeq 1/3$. Note that the value of $\chi$ found here is specific to the sinusoidal shear setup used, because both $\hat u'_{\rm rms}$ and $\hat w_{\rm rms}$ are derived from whole-domain averages, while the shear is not constant within the domain. In other words, using a different model setup such as a shearing box could yield a different value of $\chi$, though we expect it to remain close to unity. 

With this information, we can now compute the total velocity $\hat v_{tf}$ within a tilted finger (as illustrated in Figure \ref{fig:tilt}): 
\begin{equation}
\hat v_{tf} = \sqrt{ \hat u_{tf}^2 + \hat w_{tf}^2} =\hat w_{tf}  \sqrt{ 1 + \chi^2 \left(\frac{\hat S}{ \hat \lambda_f}\right)^2  } .
\end{equation} 
As in the unsheared case described earlier, we now assume that saturation of the instability occurs when the growth rate of parasitic shear instabilities between two adjacent fingers is of the order of the growth rate of the fingering instability itself. In the tilted finger $\hat \sigma = \eta \hat v_{tf}  \hat l_{tf}$ where $\hat l_{tf}$ is the wavenumber of tilted finger, and $\eta$ is the same universal constant as before. Using simple geometrical arguments (see Figure \ref{fig:tilt}), it is relatively easy to show that 
\begin{equation}
\hat l_{tf} = \hat l_{f}  \sqrt{ 1 + \chi^2 \left(\frac{\hat S}{ \hat \lambda_f}\right)^2  } 
\end{equation}
so we can expect saturation when 
\begin{equation}
\hat \lambda_f  = C_B \eta  \hat l_{tf}  \hat v_{tf} = C_B \eta \hat l_{f}  \hat w_{tf}  \left( 1 + \chi^2 \left(\frac{\hat S}{ \hat \lambda_f}\right)^2  \right),
\end{equation}
which implies 
\begin{equation}
 \hat w_{tf}  \simeq \frac{ \hat \lambda_f }{C_B \eta  \hat l_{f}  } \left( 1 + \chi^2 \left(\frac{\hat S}{ \hat \lambda_f}\right)^2  \right)^{-1} =  \hat w_{f}  \left( 1 + \chi^2 \left(\frac{\hat S}{ \hat \lambda_f}\right)^2  \right)^{-1},
\label{eq:wtf}
\end{equation}
using equation (\ref{eq:parasitic}). With this, we can compute the expected compositional perturbation within the tilted finger $\hat C_{tf}$, using the modal equation for the evolution of the compositional perturbations:  
\begin{equation}
\hat \lambda_f \hat C_{tf} + R_0^{-1} \hat w_{tf} = -\tau \hat l_f^2  \hat C_{tf}  \rightarrow \hat C_{tf} = - \frac{R_0^{-1} \hat w_{tf}  }{  \hat \lambda_f   + \tau \hat l_f^2} .
\end{equation}
Note that we have used $\hat l_f$ instead of $\hat l_{tf}$ in this expression, which is done for simplicity (see below); the results are not strongly affected by this simplification, since the compositional diffusion term is usually very small in stars anyway ($\tau \ll 1$). 
Finally, we find that the magnitude of the compositional flux in tilted fingers is given by 
\begin{eqnarray}
\hat F_C(\hat S) = K_B  \hat w_{tf}   \hat C_{tf}  = K_B \frac{R_0^{-1} \hat w^2_{tf}  }{\hat  \lambda_f    + \tau \hat l_f^2}  = K_B \frac{R_0^{-1} \hat w^2_{f}  }{ \hat \lambda_f   + \tau \hat l_f^2}\left( 1 + \chi^2 \left(\frac{\hat S}{ \hat \lambda_f}\right)^2  \right)^{-2} \nonumber \\
= \hat F_C(\hat S = 0)   \left( 1 + \chi^2 \left(\frac{\hat S}{ \hat \lambda_f}\right)^2  \right)^{-2} , 
\label{eq:fluxmod}
\end{eqnarray}
using equations (\ref{eq:brownflux}) and (\ref{eq:wtf}). Note how the ratio $\hat F_C / \hat F_C(0)$ only depends on $\hat S / \hat \lambda_f$ in this model, a result that is reminiscent of the linear stability analysis performed in Section \ref{sec:linres}. 

\begin{figure}[h]
\epsscale{0.5}
\plotone{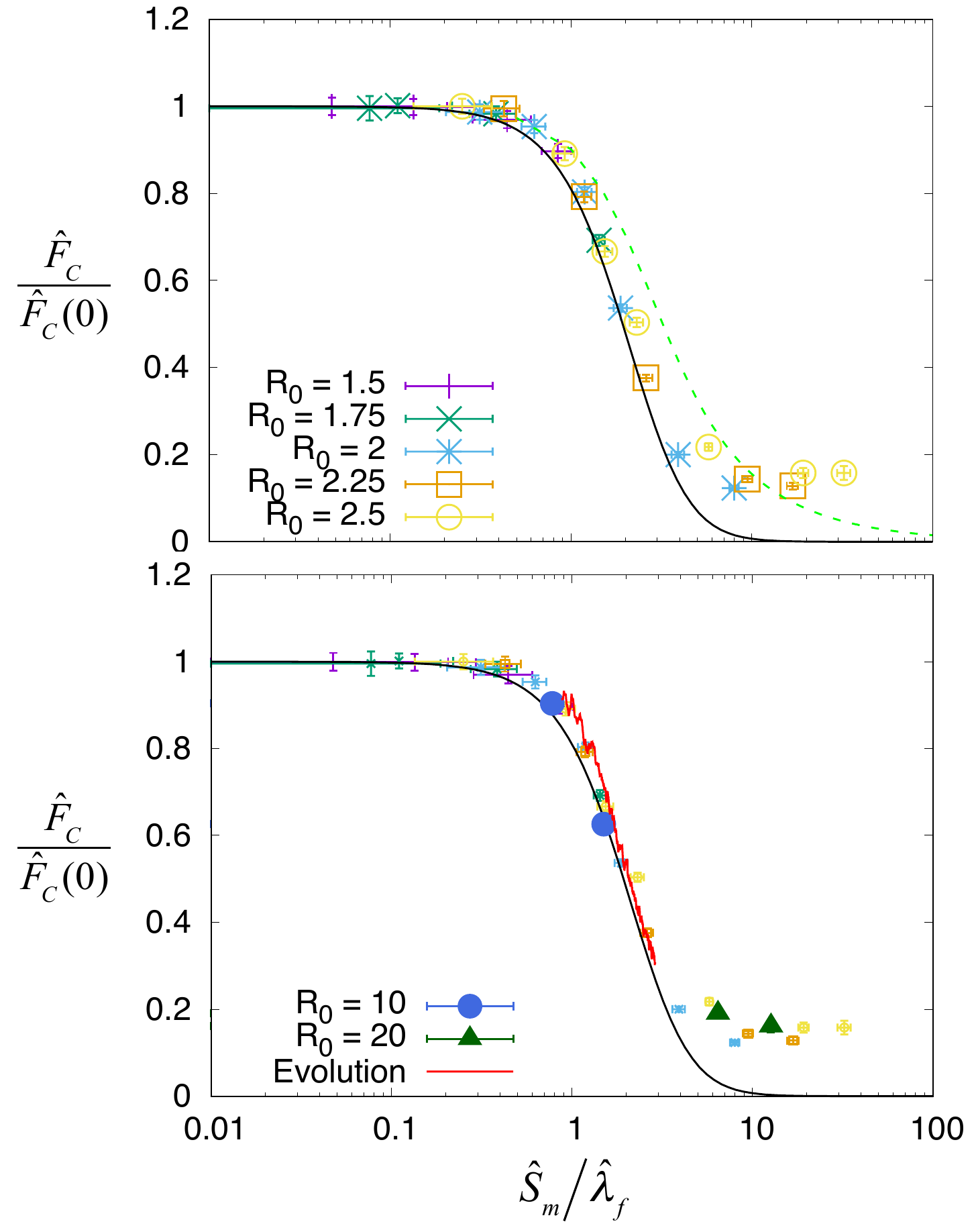}
\caption{Top: Variation of $\hat F_C / \hat F_C(0)$ against $\hat S_m / \hat \lambda_f$, where all the quantities are defined in the main text, for all the simulations reported in Table \ref{tab:data1}.  Different symbols represent different values of $R_0$, as described in the legend. The black line is the prediction from equation (\ref{eq:fluxmod}), and the green dotted line is from (\ref{eq:betterflux}). Bottom: As above for the small symbols, but now including additional simulations. The large filled-in symbols correspond to the Table \ref{tab:data2} simulations at ${\rm Pr} = \tau = 0.03$, for two different values of $R_0$ (see legend). The red line shows the temporal evolution of a simulation with ${\rm Pr} = \tau = 0.3$, $R_0 = 1.5$, and ${\rm Ri} = 0.01$ (see text for detail)}
\label{fig:FluxvsSoverl}
\end{figure}

This prediction can be compared with the data from our DNSs, but the comparison is complicated by the fact that the shear is not constant in the domain, which has a number of consequences described below. If we ignore the problem for now, and simply compare the ratio $\hat F_C / \hat F_C(0)$ of the volume-averaged  compositional flux measured in sheared simulations to the corresponding flux from unsheared simulations, with $\hat S_m / \hat \lambda_f$ (where $\hat S_m$ is the maximum amplitude of the shear measured from the mean flow, see above), we obtain Figure \ref{fig:FluxvsSoverl}a. In all cases, $\hat F_C  = \tau R_0 \langle | \nabla \hat C |^2  \rangle $  (see equation \ref{eq:FCdef}) is computed once the system has achieved a statistically stationary state, while $\hat F_C(0)$ is similarly measured from a simulation at the same values of $R_0$, Pr, and $\tau$, but without shear. We see that the data collapses onto a single universal curve, demonstrating that the relative effect of shear depends only on $\hat S_m / \hat \lambda_f$, as in the model (\ref{eq:fluxmod}). The solid black line is the prediction from (\ref{eq:fluxmod}) with $\chi = 1/3$, and we see that it fits the data very well up to $\hat S_m / \hat \lambda_f \simeq 3$. Note that no additional constant was fitted to create Figure \ref{fig:FluxvsSoverl}a, since the value of $\chi = 1/3$ had already  been independently fitted earlier to the data relating $\hat u'_{\rm rms}/\hat w_{\rm rms}$ to $\hat S_m / \hat \lambda_f \simeq 3$ in Figure \ref{fig:urmswrms}. 

Beyond $\hat S_m / \hat \lambda_f \simeq 3$, we see that the model seriously underestimates transport, and goes to 0 as $(\hat S_m / \hat \lambda_f)^{-4}$, while the data reaches a plateau with $\hat F_C / \hat F_C(0) \simeq 0.15$. The origin of this discrepancy remains to be determined and may (or may not) depend on the model setup used in this paper. Several possibilities come to mind, but none of them satisfactorily solve the problem. 

The most obvious explanation for this discrepancy is the fact that we neglected the variation of $\hat S$ with height in the domain when producing Figure \ref{fig:FluxvsSoverl}a, and effectively assumed that $\hat S = \hat S_m$ everywhere. This overestimates the amount of shear, and therefore underestimates the averaged turbulent compositional flux (which is larger in regions of weaker shear). To compute a ``better" estimate of the compositional flux integrated over the computational domain in our sinusoidally forced simulations, we can approximate $\hat S(z) \simeq \hat S_m \cos(\hat k_s z)$, and evaluate the integral 
\begin{equation} 
\frac{\hat F_C}{\hat F_C(\hat S = 0) }  = \frac{1}{L_z} \int_0^{L_z}    \left( 1 + \chi^2 \left(\frac{\hat S(z)}{ \hat \lambda_f}\right)^2  \right)^{-2}  dz = \frac{1 }{2\pi}    \int_0^{2\pi}  \left( 1 + \chi^2 \frac{\hat S_m^2}{\hat \lambda_f^2} \cos^2( \zeta)    \right)^{-2}  d\zeta . 
\label{eq:betterflux}
\end{equation}
We computed the integral numerically as a function of $\hat S_m / \hat \lambda_f$ (with $\chi = 1/3$) and the results are shown in the green dashed line on Figure \ref{fig:FluxvsSoverl}a. We see that, as expected from the discussion above, $\hat F_C$ is larger than in the case where we assumed $\hat S = \hat S_m$  everywhere, but still goes to 0 as $\hat S_m/ \hat \lambda_f$ increases, and fails to explain the plateau in the data. This is not entirely surprising, however. In order to properly account for the effect of a varying shear, we would also need to account for the fact that the vertically varying compositional flux implies a vertically varying background compositional gradient, so the local density ratio is no longer constant in the domain. As a result, the growth rate of the fingers will also depend on $z$, and the model given in (\ref{eq:fluxmod}) needs to be further modified to include that effect. For these reasons, it appears that a better way of testing (\ref{eq:fluxmod}) in the limit of strong shear would be to use simulations in a shearing box model, where the background shear is held constant by construction; this will be the subject of future work. 

Other explanations for the enhanced transport at high shearing rates are also possible. For instance, the shear itself could become unstable to {\it diffusive} shear instabilities on the finger scale. While we have checked that standard shear instabilities are never excited in the simulations presented because the gradient Richardson number $J = \hat N^2 / \hat S_m^2 = {\rm Pr} (1-R_0^{-1})/{\hat S_m^2}$ is always significantly larger than one, diffusive shear instabilities could take place, as explained in Section \ref{sec:model} \citep[see, e.g.][]{GaraudKulen16,GaraudGagnier2017}. These are difficult to identify unambiguously, however, since they are subcritical in nature, and it is not clear how to distinguish them beyond doubt from other dynamics. The fact that $\hat \nu_{\rm turb} \simeq \hat \kappa_{C, {\rm turb}}$ in that limit (see Figure \ref{fig:DnuvsDC} and associated text at the end of Section \ref{sec:num}) could be a clue to their presence, since this scaling seems to be an intrinsic property of diffusive shear-induced turbulence \citep{Pratal2016,GaraudGagnier2017}. However, if that was indeed the case, it is not clear why the compositional flux $\hat F_C$ should be equal to a constant fraction of $\hat F_C(0)$ in that limit. Simulations in a different model setup should first be performed to determine whether they also have $\hat F_C / \hat F_C(0) \simeq  0.15$ for  $\hat S_m / \hat \lambda_f > 3$, before proceeding with constructing theories for this regime.   

Finally note that for even larger values of $\hat S_m / \hat \lambda_f$ (not shown here), $\hat F_C / \hat F_C(0)$ increases again once the shear instability takes over the fingering instability. This happens when the gradient Richardson number $J$ drops below a critical threshold, at which point the nature of the perturbations changes entirely. Studying this regime is much more computationally demanding, however, because it requires a domain that is at a minimum the same size as we presently use in the $x-$ and $z-$ directions, but much wider in the $y-$direction, to capture the three-dimensionality of the much larger-scale shear-induced eddies. This may be discussed in a future publication.

Having compared our model for compositional transport by sheared fingering convection against data from simulations that had reached a statistically stationnary state, we may now wonder whether that model also applies when the system is slowly evolving in time, a situation that is more likely to be realized in stellar interiors. Figure \ref{fig:FluxvsSoverl}b shows that it does. The red line shows the temporal evolution of $\hat F_C(\hat S(t))/ \hat F_C(0)$ against $\hat S_m(t) / \hat \lambda_f$, in a run with $R_0 = 1.5$ and ${\rm Ri} = 0.01$ that was initialized from the statistically stationary state reached by the $R_0 = 1.5$, ${\rm Ri} = 0.1$ simulation described in Table \ref{tab:data1} (with the same resolution and domain size). The increase in the forcing associated with the reduction of ${\rm Ri}$ causes the mean flow to gradually accelerate, and the mean shear to increase with time. As a result, the compositional flux gradually decreases. We see that the flux law (\ref{eq:fluxmod}) derived above still provides a good estimate for transport in this system at each point in time. 

Similarly, to see whether the model applies for other values of the Prandtl number and diffusivity ratio, we have run 4 additional simulations for ${\rm Pr} = \tau =  0.03$ (see Table \ref{tab:data2}). Two of these have a density ratio of $R_0 = 10$, which is relatively low at these parameters, and two have a density ratio of $R_0 = 20$, which is closer to marginal stability. Despite the large differences in the governing parameters, we see that these simulations lie close to the existing data at higher ${\rm Pr}$ and $\tau$, for {\it all} values of $\hat S_m / \hat \lambda_f$ (smaller and greater than 3). This suggests that, at smaller values of ${\rm Pr}$ and $\tau$, (\ref{eq:fluxmod}) continues to hold for low shearing rates, and that $\hat F_C / \hat F_C(0) \simeq 0.15 $ continues to be true as well for larger shearing rates. We also see in Figure \ref{fig:DnuvsDC} that $\hat \nu_{\rm turb} \simeq 0.25 \hat \kappa_{C,{\rm turb}}$ for the fingering-dominated $R_0 = 10$ simulations and that $\hat \nu_{\rm turb} \simeq \hat \kappa_{C,{\rm turb}}$ for the shear-dominated $R_0 = 20$ simulations, so these results also appear to be robust regardless of Pr and $\tau$. 
 
 \begin{table}[]
        \caption{\small As for Table \ref{tab:data1}, but for ${\rm Pr} = \tau = 0.03$. These simulations were run in a shorter domain size, with $L_x = 200$ ($L_y  =25$ and $L_z = 200$ remain unchanged), and have an effective resolution of ($384 \times 48 \times 384$) equivalent grid points. A statistically stationary state was achieved in all cases. }
        \label{table1}
        \centering{
        \vspace{0.3cm}
        {\small 
\begin{tabular}{cccccccc}
                \tableline
                 $R_0$ & ${\rm Ri}$ &  $\hat u_{\rm rms}$        &       $\hat w_{\rm rms}$     &     max$\left(\overline{\hat u}\right)$  &   $\hat S_m$ &         $ \langle |\nabla \hat C|^2 \rangle $      & $\hat \nu_{\rm turb}$           \\
                        \tableline
  10 & 1 & 0.66 $\pm$ 0.003 & 0.11 $\pm$ 0.001 & 0.95  $\pm$   0.01 & 0.027 $\pm$  0.005 & 0.144 $\pm$ 0.003 & 0.144 $\pm$ 0.01 \\ 
  10 & 3 & 0.328 $\pm$ 0.001 & 0.136  $\pm$ 0.001 & 0.45  $\pm$   0.01 & 0.015 $\pm$  0.005 & 0.207  $\pm$ 0.003 & 0.17 $\pm$ 0.01 \\ 
  20 & 3 & 1.299 $\pm$ 0.002 & 0.015 $\pm$ 0.0006 & 1.84  $\pm$   0.02 & 0.057 $\pm$  0.004 & 0.002  $\pm$ 0.0001 & 0.021 $\pm$ 0.002 \\ 
  20 & 10 & 0.683 $\pm$ 0.001 & 0.017  $\pm$ 0.0005 & 0.97  $\pm$   0.01 & 0.029 $\pm$  0.003 & 0.0023  $\pm$ 0.0001 & 0.024 [0.015,0.027] \\ 
 \\
\tableline
\end{tabular}
}
}
\label{tab:data2}
\end{table}

\section{Implications for stellar interiors, and future work}
\label{sec:ccl}

Despite the uncertainties remaining in the theory described above, we have established that the effect of shear on fingering convection depends on a single quantity, namely the ratio of the shearing rate to the intrinsic (unsheared) finger growth rate $\hat S / \hat \lambda_f$, or, in terms of dimensional quantities, $S / \lambda_f$. We have also established that moderate shear reduces the efficiency of fingering, and that the effect becomes pronounced when $S /  \lambda_f$ is of order unity, or equivalently, when the shearing rate approaches the fingering growth rate $\lambda_f$. The formula for the compositional flux given in equation (\ref{eq:fluxmod}) can be re-written in terms of a turbulent mixing coefficient as 
\begin{equation}
 \kappa_{C,{\rm turb}}(S) =   \kappa_{C,{\rm turb}}  (S = 0)   \left( 1 + \chi^2 \left(\frac{S}{ \lambda_f}\right)^2  \right)^{-2} , 
\end{equation}
where $\chi = 1/3$ for sinusoidal shear flows (though could be a little different in a different setup), and  $\kappa_{C,{\rm turb}}  (S = 0)$ is the turbulent mixing coefficient in the absence of shear, which can be computed for instance using the model of \citet{Brownal2013}. The formula appears to be valid up to $S / \lambda_f \simeq 3$, beyond which it underestimates  $\kappa_{C,{\rm turb}}$. For  $S / \lambda_f > 3$, we find that  $\kappa_{C,{\rm turb}}(S) \simeq 0.15  \kappa_{C,{\rm turb}}  (S = 0)$ instead, although this result needs to be confirmed using a different model setup (such as a shearing box, for instance).

We have also studied momentum transport by fingering convection, and have found that the turbulent viscosity and the turbulent mixing coefficient are indeed (almost) proportional to one another, with $\nu_{\rm turb} \simeq 0.25 \kappa_{C,{\rm turb}}$ when $S/ \lambda_f < 3$, gradually increasing to $\nu_{\rm turb} \simeq \kappa_{C,{\rm turb}}$ when $S/ \lambda_f > 3$. This higher-shear limit remains to be confirmed using a different model setup, however, as discussed in Section \ref{sec:simplemodel}.

Note that all of these results have been obtained for moderate shear, prior to the onset of standard shear instabilities (i.e. when $J = N^2 / S^2$ is larger than one). For larger shearing rates, we expect shear instabilities to develop, although this limit has not been investigated yet. 

Using this information, we can conjecture on the conditions where shear might substantially reduce the fingering-induced mixing efficiency in stellar interiors. To estimate the dimensional finger growth rate, we use the asymptotic formula of \citet{Brownal2013} (see their Appendix B.2) for $1 \ll R_0 \ll \tau^{-1}$ (which is the more likely scenario in stars where $\tau$ is asymptotically small) namely 
\begin{equation}
\lambda_f = \hat \lambda_f \frac{\kappa_T}{d^2} \simeq \sqrt{ \frac{\rm Pr}{R_0}} \frac{\kappa_T}{d^2} = \frac{N_T}{\sqrt{R_0} }, 
\end{equation}
where $N^2_T = \alpha g (T_{0z} - T_{{\rm ad},z})$ is the Brunt-V\"ais\"al\"a frequency associated with the temperature stratification only (see Section \ref{sec:model}). With this estimate, we predict that shear may become relevant when $\lambda_f \sim S$, which is 
\begin{equation}
 S \sim \frac{N_T}{\sqrt{R_0} }   \rightarrow  J_T \equiv \frac{N_T^2}{S^2} \sim R_0.
\end{equation} 
In other words, fingering becomes affected by shear when the gradient Richardson number associated with the temperature stratification only is of the order of (or lower than) the density ratio $R_0$.  Such conditions, while not necessarily ubiquitous, are nevertheless possible depending on the current phase of evolution and conditions within the star. 

In Red Giant Branch stars for instance, a fingering region induced by nuclear reactions that transform $^3$He into $^4$He is located between the hydrogen burning shell and the convective envelope \citep{Ulrich1972}, and is thought to contribute to the transport of both Lithium and CNO cycle by-products between the two regions \citep{CharbonnelZahn2007}. Using the estimates from \citet{denissenkov2010} for typical parameters appropriate of that region, we have $N_T^2 \simeq 2 \times 10^{-4}$s$^{-2}$, and $R_0 \simeq 2 \times 10^3$. The local shear on the other hand is unknown, the best available constraints coming from asteroseismic estimates of the core and envelope angular velocities $\Omega_{\rm core} / 2\pi \simeq 10^{-6}$Hz and $\Omega_{\rm env}/2\pi \simeq 10^{-7}$Hz  \citep{Deheuvels2014}. If we assume the transition takes place smoothly over the entire core (from the center to the base of the convective envelope), then $S \simeq \Delta \Omega \simeq 2\pi \times 10^{-6}$s$^{-1}$, which would imply that $J_T \simeq 5 \times 10^6$. Since this is much larger than $R_0$, we conclude that a smooth shear profile in these objects would not have any effect on the fingering efficiency. On the other hand if the transition between the core and envelope rotation rates occurs over a distance $\Delta$ that is a fraction of the core size $r_{\rm core}$, then 
\begin{equation}
J_T \simeq 5 \times 10^6 \left(\frac{\Delta}{r_{\rm core}}\right)^2,
\end{equation}
and so the shear could begin to affect the fingering efficiency if $\Delta/ r_{\rm core} \simeq 0.01$ or less\footnote{unless $R_0$ is much larger than estimated by \citet{denissenkov2010}}, which seems unlikely.  As a result, we predict that the turbulent mixing coefficient $\kappa_{C,{\rm turb}}$ in Red Giant Branch stars (ignoring magnetic fields) can be estimated using the model proposed by \citet{Brownal2013} in the absence of shear, and further find that the turbulent viscosity can be computed as $\nu_{\rm turb} \simeq 0.25 \kappa_{C,{\rm turb}}$. If magnetic fields are present, however, then the model of  \citet{HarringtonGaraud2019} for $\kappa_{C,{\rm turb}}$ should be used instead, although no model presently exists for $\nu_{\rm turb}$ in that case. 

In White Dwarfs undergoing accretion of material from a surrounding debris disk, a fingering region is thought to be located just beneath the surface \citep[c.f.][for instance]{Deal2013,BauerBildsten2018}, and could participate in draining the excess metallicity of accreted material inward in addition to the effect of gravitational settling. Assuming a steady-state balance between the rate of debris accretion and the inward subsurface flux of heavy elements, one can in theory predict their expected surface abundances. Conversedly, the observed abundances can in principle be used to infer the accretion rate of debris assuming the subsurface flux is known. Unfortunately this inference is complicated by the fact that the fingering flux depends on many different processes, as discussed in Section \ref{sec:intro}, which have all been neglected to date. To determine whether shear in particular could substantially reduce the efficiency of fingering convection in White Dwarfs, one would need to know how both the gradient Richardson number $J_T$ and the density ratio $R_0$ evolve with time after each accretion event. Qualitatively speaking, we anticipate that the density ratio $R_0$ should start from a value close to unity at the onset of the accretion event (where the inverse $\mu$ gradient is the largest), and should then gradually increase towards marginal stability ($R_0 = 1/\tau$) as the inward transport of heavy elements proceeds. Accretion is also expected to generate substantial shear (as the accreting material lands on the star at close to Keplerian speed), which would then gradually decrease as a result of the turbulent momentum transport. Because of these processes, both $J_T$ and $R_0$ are likely to increase with time after accretion, so whether shear is ever important ($J_T < R_0$) or not will depend on the initial angular momentum and composition of the accreting material relative to that of the star, as well as the efficiency of both compositional and momentum transport. This will need to be established on a case-by-case basis for each star and each accretion scenario considered. Finding substantial reduction of the fingering efficiency by the shear would be an interesting outcome of this exercise, however, since it could more easily help interpret the observed heavy-element abundances of polluted DA White Dwarfs, that can otherwise only be explained by very large accretion rates \citep{BauerBildsten2018}.

\acknowledgments

 The simulations were performed using the PADDI code kindly provided by S. Stellmach, on the Hyades supercomputer purchased using an NSF MRI grant. P. G. and J. S. acknowledge funding by NSF-AAG 1517927. A.K. is supported by the Dean's fellowship and Regents' fellowship from the Baskin School of Engineering at UC Santa Cruz.

\bibliographystyle{aasjournal}
\bibliography{DDClowPr_references}

\begin{thebibliography}{}
\expandafter\ifx\csname natexlab\endcsname\relax\def\natexlab#1{#1}\fi
\providecommand{\url}[1]{\href{#1}{#1}}

\bibitem[{Baines \& Gill(1969)}]{baines1969}
Baines, P., \& Gill, A. 1969, J. Fluid Mech., 37, 289

\bibitem[{Balmforth \& Young(2002)}]{balmforthyoung2002}
Balmforth, N.~J., \& Young, Y.-N. 2002, Journal of Fluid Mechanics, 450,
  131?167

\bibitem[{Balmforth \& Young(2005)}]{balmforthyoung2005}
---. 2005, Journal of Fluid Mechanics, 528, 23?42

\bibitem[{{Bauer} \& {Bildsten}(2018)}]{BauerBildsten2018}
{Bauer}, E.~B., \& {Bildsten}, L. 2018, \apjl, 859, L19

\bibitem[{{Beck} {et~al.}(2012){Beck}, {Montalban}, {Kallinger}, {De Ridder},
  {Aerts}, {Garc{\'{\i}}a}, {Hekker}, {Dupret}, {Mosser}, {Eggenberger},
  {Stello}, {Elsworth}, {Frandsen}, {Carrier}, {Hillen}, {Gruberbauer},
  {Christensen-Dalsgaard}, {Miglio}, {Valentini}, {Bedding}, {Kjeldsen},
  {Girouard}, {Hall}, \& {Ibrahim}}]{Beck2012}
{Beck}, P.~G., {Montalban}, J., {Kallinger}, T., {et~al.} 2012, \nat, 481, 55

\bibitem[{{Brown} {et~al.}(2013){Brown}, {Garaud}, \&
  {Stellmach}}]{Brownal2013}
{Brown}, J.~M., {Garaud}, P., \& {Stellmach}, S. 2013, Astrophys. J., 768, 34

\bibitem[{{Brown} {et~al.}(1989){Brown}, {Christensen-Dalsgaard},
  {Dziembowski}, {Goode}, {Gough}, \& {Morrow}}]{Brownal1989}
{Brown}, T.~M., {Christensen-Dalsgaard}, J., {Dziembowski}, W.~A., {et~al.}
  1989, \apj, 343, 526

\bibitem[{{Charbonneau} {et~al.}(1999){Charbonneau}, {Christensen-Dalsgaard},
  {Henning}, {Larsen}, {Schou}, {Thompson}, \& {Tomczyk}}]{Charbonneaual1999}
{Charbonneau}, P., {Christensen-Dalsgaard}, J., {Henning}, R., {et~al.} 1999,
  \apj, 527, 445

\bibitem[{{Charbonnel} \& {Zahn}(2007)}]{CharbonnelZahn2007}
{Charbonnel}, C., \& {Zahn}, J.-P. 2007, Astron. Astrophys., 467, L15

\bibitem[{{Chen} \& {Han}(2004)}]{ChenHan2004}
{Chen}, X., \& {Han}, Z. 2004, \mnras, 355, 1182

\bibitem[{{Christensen-Dalsgaard} \& {Schou}(1988)}]{JCDSchou88}
{Christensen-Dalsgaard}, J., \& {Schou}, J. 1988, in ESA Special Publication,
  Vol. 286, Seismology of the Sun and Sun-Like Stars, ed. E.~J. {Rolfe}

\bibitem[{{Deal} {et~al.}(2013){Deal}, {Deheuvels}, {Vauclair}, {Vauclair}, \&
  {Wachlin}}]{Deal2013}
{Deal}, M., {Deheuvels}, S., {Vauclair}, G., {Vauclair}, S., \& {Wachlin},
  F.~C. 2013, Astron. Astrophys., 557, L12

\bibitem[{{Deal} {et~al.}(2016){Deal}, {Richard}, \& {Vauclair}}]{Dealal2016}
{Deal}, M., {Richard}, O., \& {Vauclair}, S. 2016, \aap, 589, A140

\bibitem[{{Deheuvels} {et~al.}(2014){Deheuvels}, {Do{\u g}an}, {Goupil},
  {Appourchaux}, {Benomar}, {Bruntt}, {Campante}, {Casagrande}, {Ceillier},
  {Davies}, {De Cat}, {Fu}, {Garc{\'{\i}}a}, {Lobel}, {Mosser}, {Reese},
  {Regulo}, {Schou}, {Stahn}, {Thygesen}, {Yang}, {Chaplin},
  {Christensen-Dalsgaard}, {Eggenberger}, {Gizon}, {Mathis},
  {Molenda-{\.Z}akowicz}, \& {Pinsonneault}}]{Deheuvels2014}
{Deheuvels}, S., {Do{\u g}an}, G., {Goupil}, M.~J., {et~al.} 2014, \aap, 564,
  A27

\bibitem[{{Denissenkov}(2010)}]{denissenkov2010}
{Denissenkov}, P.~A. 2010, Astrophys. J., 723, 563

\bibitem[{{Denissenkov} \& {Merryfield}(2011)}]{DenissenkovMerryfield2011}
{Denissenkov}, P.~A., \& {Merryfield}, W.~J. 2011, Astrophys. J. Lett., 727, L8

\bibitem[{{Gagnier} \& {Garaud}(2018)}]{GagnierGaraud2018}
{Gagnier}, D., \& {Garaud}, P. 2018, \apj, 862, 36

\bibitem[{{Garaud}(2011)}]{Garaud2011}
{Garaud}, P. 2011, Astrophys. J. Lett., 728, L30

\bibitem[{Garaud(2018)}]{Garaud18}
Garaud, P. 2018, Annual Review of Fluid Mechanics, 50, 275

\bibitem[{{Garaud} \& {Brummell}(2015)}]{GaraudBrummell2015}
{Garaud}, P., \& {Brummell}, N. 2015, Astrophys. J., 815, 42

\bibitem[{{Garaud} {et~al.}(2017){Garaud}, {Gagnier}, \&
  {Verhoeven}}]{GaraudGagnier2017}
{Garaud}, P., {Gagnier}, D., \& {Verhoeven}, J. 2017, \apj, 837, 133

\bibitem[{{Garaud} {et~al.}(2015{\natexlab{a}}){Garaud}, {Gallet}, \&
  {Bischoff}}]{Garaudal15a}
{Garaud}, P., {Gallet}, B., \& {Bischoff}, T. 2015{\natexlab{a}}, Physics of
  Fluids, 27, 084104

\bibitem[{{Garaud} \& {Kulenthirarajah}(2016)}]{GaraudKulen16}
{Garaud}, P., \& {Kulenthirarajah}, L. 2016, \apj, 821, 49

\bibitem[{{Garaud} {et~al.}(2015{\natexlab{b}}){Garaud}, {Medrano}, {Brown},
  {Mankovich}, \& {Moore}}]{Garaudal2015}
{Garaud}, P., {Medrano}, M., {Brown}, J.~M., {Mankovich}, C., \& {Moore}, K.
  2015{\natexlab{b}}, Astrophys. J., 808, 89

\bibitem[{{Harrington} \& {Garaud}(2019)}]{HarringtonGaraud2019}
{Harrington}, P.~Z., \& {Garaud}, P. 2019, \apjl, 870, L5

\bibitem[{{Holyer}(1983)}]{Holyer1983}
{Holyer}, J.~Y. 1983, J. Fluid Mech., 137, 347

\bibitem[{{Holyer}(1984)}]{Holyer1984}
---. 1984, Journal of Fluid Mechanics, 147, 169

\bibitem[{Howard(1961)}]{howard1961}
Howard, L.~N. 1961, Journal of Fluid Mechanics, 10, 509

\bibitem[{Kimura \& Smyth(2007)}]{KimuraSmyth2007}
Kimura, S., \& Smyth, W. 2007, Geophysical Research Letters, 34,
  doi:10.1029/2007GL031935

\bibitem[{{Kippenhahn} {et~al.}(1980){Kippenhahn}, {Ruschenplatt}, \&
  {Thomas}}]{kippenhahn80}
{Kippenhahn}, R., {Ruschenplatt}, G., \& {Thomas}, H. 1980, Astron. Astrophys.,
  91, 175

\bibitem[{Kippenhahn {et~al.}(1990)Kippenhahn, Weigert, \&
  Weiss}]{kippenhahnweigert}
Kippenhahn, R., Weigert, A., \& Weiss, A. 1990, Stellar structure and
  evolution, Vol. 192 (Springer)

\bibitem[{{Kumar} {et~al.}(1999){Kumar}, {Talon}, \& {Zahn}}]{Kumaral1999}
{Kumar}, P., {Talon}, S., \& {Zahn}, J.-P. 1999, \apj, 520, 859

\bibitem[{Kunze(1990)}]{kunze1990}
Kunze, E. 1990, Journal of Marine Research, 48, 471

\bibitem[{Kunze(1994)}]{kunze1994}
---. 1994, Journal of marine research, 52, 999

\bibitem[{{Linden}(1974)}]{Linden1974}
{Linden}, P.~F. 1974, Geophysical and Astrophysical Fluid Dynamics, 6, 1

\bibitem[{{Malkus}(1954)}]{Malkus1954}
{Malkus}, W.~V.~R. 1954, Proceedings of the Royal Society of London Series A,
  225, 196

\bibitem[{{Marks} \& {Sarna}(1998)}]{MarksSarna1998}
{Marks}, P.~B., \& {Sarna}, M.~J. 1998, \mnras, 301, 699

\bibitem[{{Marques} {et~al.}(2013){Marques}, {Goupil}, {Lebreton}, {Talon},
  {Palacios}, {Belkacem}, {Ouazzani}, {Mosser}, {Moya}, {Morel}, {Pichon},
  {Mathis}, {Zahn}, {Turck-Chi{\`e}ze}, \& {Nghiem}}]{Marques2013}
{Marques}, J.~P., {Goupil}, M.~J., {Lebreton}, Y., {et~al.} 2013, Astron. \&
  Astrophys., 549, A74

\bibitem[{{Medrano} {et~al.}(2014){Medrano}, {Garaud}, \&
  {Stellmach}}]{Medrano2014}
{Medrano}, M., {Garaud}, P., \& {Stellmach}, S. 2014, Astrophys. J. Lett., 792,
  L30

\bibitem[{Miles(1961)}]{miles1961}
Miles, J.~W. 1961, Journal of Fluid Mechanics, 10, 496

\bibitem[{{Mirouh} {et~al.}(2012){Mirouh}, {Garaud}, {Stellmach}, {Traxler}, \&
  {Wood}}]{Mirouh2012}
{Mirouh}, G.~M., {Garaud}, P., {Stellmach}, S., {Traxler}, A.~L., \& {Wood},
  T.~S. 2012, Astrophys. J., 750, 61

\bibitem[{{Ogilvie}(2014)}]{Ogilvie2014}
{Ogilvie}, G.~I. 2014, \araa, 52, 171

\bibitem[{{Paparella} \& {Spiegel}(1999)}]{PaparellaSpiegel1999}
{Paparella}, F., \& {Spiegel}, E.~A. 1999, Physics of Fluids, 11, 1161

\bibitem[{{Paxton} {et~al.}(2011){Paxton}, {Bildsten}, {Dotter}, {Herwig},
  {Lesaffre}, \& {Timmes}}]{MESA12011}
{Paxton}, B., {Bildsten}, L., {Dotter}, A., {et~al.} 2011, \apjs, 192, 3

\bibitem[{{Prat} {et~al.}(2016){Prat}, {Guilet}, {Viallet}, \&
  {M{\"u}ller}}]{Pratal2016}
{Prat}, V., {Guilet}, J., {Viallet}, M., \& {M{\"u}ller}, E. 2016, Astron. \&
  Astrophys., 592, A59

\bibitem[{{Prat} \& {Ligni{\`e}res}(2013)}]{PratLignieres13}
{Prat}, V., \& {Ligni{\`e}res}, F. 2013, Astron. \& Astrophys., 551, L3

\bibitem[{{Prat} \& {Ligni{\`e}res}(2014)}]{PratLignieres14}
---. 2014, Astron. \& Astrophys., 566, A110

\bibitem[{Radko {et~al.}(2015)Radko, Ball, Colosi, \& Flanagan}]{Radkoal2015}
Radko, T., Ball, J., Colosi, J., \& Flanagan, J. 2015, Journal of Physical
  Oceanography, 45, 3155

\bibitem[{{Radko} \& {Smith}(2012)}]{RadkoSmith2012}
{Radko}, T., \& {Smith}, D.~P. 2012, J. Fluid Mech., 692, 5

\bibitem[{{Rosenblum} {et~al.}(2011){Rosenblum}, {Garaud}, {Traxler}, \&
  {Stellmach}}]{rosenblumal2011}
{Rosenblum}, E., {Garaud}, P., {Traxler}, A., \& {Stellmach}, S. 2011,
  Astrophys. J., 731, 66

\bibitem[{Ruddick(1985)}]{ruddick1985}
Ruddick, B.~R. 1985, Journal of Geophysical Research: Oceans, 90, 895

\bibitem[{{Sengupta} \& {Garaud}(2018)}]{SenguptaGaraud2018}
{Sengupta}, S., \& {Garaud}, P. 2018, \apj, 862, 136

\bibitem[{Smyth \& Kimura(2007)}]{SmythKimura2007}
Smyth, W.~D., \& Kimura, S. 2007, Journal of Physical Oceanography, 37, 1551

\bibitem[{Smyth \& Kimura(2011)}]{SmythKimura2010}
---. 2011, Journal of Physical Oceanography, 41, 1364

\bibitem[{{Spiegel} \& {Veronis}(1960)}]{SpiegelVeronis1960}
{Spiegel}, E.~A., \& {Veronis}, G. 1960, Astrophys. J., 131, 442

\bibitem[{{Stancliffe} {et~al.}(2007){Stancliffe}, {Glebbeek}, {Izzard}, \&
  {Pols}}]{Stancliffe2007}
{Stancliffe}, R.~J., {Glebbeek}, E., {Izzard}, R.~G., \& {Pols}, O.~R. 2007,
  Astron. Astrophys., 464, L57

\bibitem[{{Stellmach} {et~al.}(2011){Stellmach}, {Traxler}, {Garaud},
  {Brummell}, \& {Radko}}]{Stellmach2011}
{Stellmach}, S., {Traxler}, A., {Garaud}, P., {Brummell}, N., \& {Radko}, T.
  2011, J. Fluid Mech., 677, 554

\bibitem[{Stern(1960)}]{stern1960sfa}
Stern, M. 1960, Tellus, 12, 172

\bibitem[{{Stommel} {et~al.}(1956){Stommel}, {Arons}, \&
  {Blanchard}}]{Stommel1956}
{Stommel}, H., {Arons}, A.~B., \& {Blanchard}, D. 1956, Deep Sea Research, 3,
  152

\bibitem[{{Th{\'e}ado} \& {Vauclair}(2012)}]{TheadoVauclair2012}
{Th{\'e}ado}, S., \& {Vauclair}, S. 2012, \apj, 744, 123

\bibitem[{{Th{\'e}ado} {et~al.}(2009){Th{\'e}ado}, {Vauclair}, {Alecian}, \&
  {LeBlanc}}]{Theadoal2009}
{Th{\'e}ado}, S., {Vauclair}, S., {Alecian}, G., \& {LeBlanc}, F. 2009, \apj,
  704, 1262

\bibitem[{{Traxler} {et~al.}(2011{\natexlab{a}}){Traxler}, {Garaud}, \&
  {Stellmach}}]{Traxler2011b}
{Traxler}, A., {Garaud}, P., \& {Stellmach}, S. 2011{\natexlab{a}}, Astrophys.
  J. Lett., 728, L29

\bibitem[{{Traxler} {et~al.}(2011{\natexlab{b}}){Traxler}, {Stellmach},
  {Garaud}, {Radko}, \& {Brummell}}]{Traxler2011a}
{Traxler}, A., {Stellmach}, S., {Garaud}, P., {Radko}, T., \& {Brummell}, N.
  2011{\natexlab{b}}, J. Fluid Mech., 677, 530

\bibitem[{{Ulrich}(1972)}]{Ulrich1972}
{Ulrich}, R.~K. 1972, Astrophys. J., 172, 165

\bibitem[{Vauclair(2004)}]{vauclair2004mfa}
Vauclair, S. 2004, Astrophys. J., 605, 874

\bibitem[{{Wachlin} {et~al.}(2017){Wachlin}, {Vauclair}, {Vauclair}, \&
  {Althaus}}]{Wachlinal2017}
{Wachlin}, F.~C., {Vauclair}, G., {Vauclair}, S., \& {Althaus}, L.~G. 2017,
  \aap, 601, A13

\bibitem[{Wells {et~al.}(2001)Wells, Griffiths, \& Turner}]{wells2001}
Wells, M.~G., Griffiths, R.~W., \& Turner, J.~S. 2001, Journal of Geophysical
  Research: Oceans, 106, 7027

\bibitem[{{Wood} {et~al.}(2013){Wood}, {Garaud}, \& {Stellmach}}]{Woodal13}
{Wood}, T.~S., {Garaud}, P., \& {Stellmach}, S. 2013, Astrophys. J., 768, 157

\bibitem[{{Xie} {et~al.}(2019){Xie}, {Julien}, \& {Knobloch}}]{Xieal2019}
{Xie}, J.-H., {Julien}, K., \& {Knobloch}, E. 2019, Journal of Fluid Mechanics,
  858, 228

\bibitem[{{Zahn}(1974)}]{Zahn1974}
{Zahn}, J.-P. 1974, in IAU Symposium, Vol.~59, Stellar Instability and
  Evolution, ed. P.~{Ledoux}, A.~{Noels}, \& A.~W. {Rodgers}, 185--194

\bibitem[{{Zahn}(1992)}]{Zahn92}
{Zahn}, J.-P. 1992, Astron. \& Astrophys., 265, 115

\bibitem[{{Zemskova} {et~al.}(2014){Zemskova}, {Garaud}, {Deal}, \&
  {Vauclair}}]{Zemskova2014}
{Zemskova}, V., {Garaud}, P., {Deal}, M., \& {Vauclair}, S. 2014, Astrophys.
  J., 795, 118

\end{thebibliography}

\end{document}